\documentclass[aps,prd,twocolumn,numerical,floatfix,nofootinbib]{revtex4-2}

\usepackage[utf8]{inputenc}
\usepackage[table,dvipsnames]{xcolor}

\usepackage{caption}
\usepackage{subcaption}
\usepackage{graphicx}

\usepackage{dcolumn}% Align table columns on decimal point
\usepackage{bm}% bold math
\usepackage{physics}
\usepackage[colorlinks=true,linkcolor=blue,citecolor=blue,urlcolor=
blue]{hyperref}
\usepackage{latexsym}

\usepackage{lipsum}
\usepackage{mwe}
\usepackage{multirow}

\newcommand{\feq}{f^{\text{eq}}}

\newcommand{\lrb}[1]{\left( #1 \right)}
\newcommand{\lrrb}[1]{\left[ #1 \right]}

\newcommand{\avg}[1]{\langle #1 \rangle }

\usepackage{ulem}

\begin{document}

	%\title{Does Electric Field Affect the Flow Harmonics in Heavy-Ion Collisions?}

	%\title{Evolution of the Moments in MDRTA}% Force line breaks with \\
	\title{On the Approach Towards Equilibrium Through Momentum-Dependent Relaxation: Insights from Evolution of the Moments in Kinetic Theory}
    % Momentum-Dependent Relaxation and Equilibration: A Kinetic Theory Perspective
    \author{Reghukrishnan Gangadharan}
    \email{reghukrishnang@niser.ac.in}
    \affiliation{School of Physical Sciences, National Institute of Science Education and Research, An OCC of Homi Bhabha National Institute, Jatni-752050, India}
	\author{Sukanya Mitra}
    \email{sukanya.mitra@niser.ac.in}
    \affiliation{School of Physical Sciences, National Institute of Science Education and Research, An OCC of Homi Bhabha National Institute, Jatni-752050, India}
    \author{Victor Roy}
    \email{victor@niser.ac.in}
    \affiliation{School of Physical Sciences, National Institute of Science Education and Research, An OCC of Homi Bhabha National Institute, Jatni-752050, India}
	\date{\today}% It is always \today, today,
\begin{abstract}
We investigate the impact of momentum-dependent relaxation time approximation in the Boltzmann equation within the Bjorken flow framework by analyzing the moments of the single-particle distribution function. The moment equations, which form an infinite hierarchy, provide important insights about the system dynamics and the approach towards equilibrium for systems far from equilibrium. The momentum-dependent collision kernel couples moments through both the energy exponents and the angular dependence via various-order Legendre polynomials, resulting in an intricate system of infinitely coupled equations. A naive truncation of the coupled equations results in diverging moments at late times. We outline strategies for solving the coupled system, including a novel approach for managing the divergences and non-integer moments. We show a significant influence of momentum dependent relaxation time on the time evolution of the moments, particularly for higher-order moments and system with smaller shear viscosity over entropy density, emphasizing the importance of incorporating such dependence for a more accurate description of the system dynamics with low shear viscosity such as the quark-gluon-plasma produced in high-energy heavy-ion collisions.

\end{abstract}
	
	%\keywords{Suggested keywords}%Use show keys class option if keyword
	\maketitle
\section{Introduction}
Over the past few decades, relativistic hydrodynamics has found widespread applications in describing the dynamical evolution of collective systems, ranging from cosmology and astrophysics to the de-confined quark-gluon plasma produced in ultra-relativistic heavy-ion collisions \cite{Baier:2007ix,Huovinen:2006jp,Gale:2013da}. This framework characterizes system evolution through conservation laws, leading to a set of coupled partial differential equations that govern macroscopic state variables and incorporate dissipative effects to account for irreversible phenomena.

Hydrodynamic theories are typically derived from an underlying microscopic description, such as covariant kinetic theory \cite{Romatschke:2009im,Denicol-Rischke-Dirk}, which encodes short-wavelength information into long-wavelength behavior, ultimately governing the system's dynamics. This connection is often established by taking moments of the relativistic transport equation \cite{DeGroot:1980dk, Cercignani1988}. While the lowest two moments yield the conservation laws for particle number and energy momentum, higher moments describe various dissipative effects within the system.

Despite the extensive success of relativistic hydrodynamics, pathologies related to its physical properties have always troubled its development. The first order relativistic Navier-Stokes (NS) theory \cite{landau_lifshitz,Eckart:1940te}  is acausal with superluminal signal propagation and unstable against small perturbations\cite{Hiscock:1983zz,Hiscock:1985z}. The higher-order theories, mostly the Muller-Israel-Stewart (MIS) theory and its other variants \cite{1967ZPhy..198..329M,Israel:1976efz,Israel:1976tn,Denicol:2012es,Muronga:2001zk} partially solve these shortcomings and are applied for practical purposes like hydrodynamic simulations of heavy-ion collisions. However, they are applicable only within a strict range of its parameter space (in terms of its transport coefficients that can be derived from the underlying microscopic theory and act as input parameters of the hydro evolution equations) \cite{Olson:1990rzl,Pu:2009fj}. Considering the scenario, a convenient alternative where these physical constraints can be bypassed could be to take suitable moments
of the Boltzmann transport equation itself and recast it into an infinite hierarchy of differential equations of moments over the single-particle momentum distribution function. The method of moments has been known to be a useful technique for quite some time \cite{Grad:1949zza,Israel:1979wp,Denicol:2012cn}. For a system where the moments converge fast enough, it suffices to consider a limited number of the first few moments in the infinite hierarchy. This technique provides an alternative yet efficient way that saves us the complexities of computing the exact solution of the Boltzmann equation.

In several recent works \cite{Blaizot:2017lht,Blaizot:2017ucy,Blaizot:2020gql,Aniceto:2024pyc}, the method of moments has been studied to provide a quantitative measure of momentum anisotropies in a dissipative medium and to investigate the onset of hydrodynamics for a longitudinally expanding boost-invariant system described by the Bjorken flow \cite{Bjorken:1982qr}. In \cite{deBrito:2024vhm,deBrito:2024qow} the convergence properties of the hierarchy of moment equations have been explored. In most of these studies, the source term in the kinetic equation (to be precise, the collision term of the relativistic Boltzmann equation) is expressed in terms of the relaxation time approximation (RTA). In its conventional form \cite{Anderson:1974nyl}, the relaxation time scale for the single-particle distribution function is taken to be independent of momenta.

Now, the momentum dependence of the relaxation time is known to be related to microscopic interactions relevant to the medium under consideration \cite{Dusling:2009df}. Following this lead, in a number of recent studies \cite{Teaney:2013gca,Kurkela:2017xis,Rocha:2021zcw,Mitra:2020gdk,Dash:2021ibx,Dash:2023ppc,Singh:2024leo,Bhadury:2024ckc,KumarSingh:2025kml,Mukherjee:2025dqp}, the momentum-dependent relaxation time approximation has been adopted to derive the relativistic hydrodynamic equations. For example, in \cite{Teaney:2013gca} Fig.[1] shows the dependence of differential scaled $v_2$ on the momentum-dependent correction to the distribution function. Considering the scenario, it is only indicative that momentum dependence should be included in the RTA approach for undertaking the moment analysis from the relativistic Boltzmann equation. The current analysis is an attempt in this direction where the sequence of moments has been derived from the Boltzmann equation using a momentum-dependent relaxation time approximation (MDRTA) with the boost invariant symmetry of Bjorken flow. The interesting features and the added complexities of this analysis are described below.

In general (irrespective of the symmetry chosen) the moments are defined as a phase-space integral over the particle distribution function weighted by a product of a power of particle energy and an irreducible tensor of particle momenta having a certain rank \cite{deBrito:2024vhm}. 
For a system that is invariant under Lorentz boosts described by Bjorken geometry \cite{Bjorken:1982qr} and consisting of massless particles (as considered in the present analysis), the definition of the moment $\rho_{n,l}$ takes the following form,
\begin{eqnarray}
    \rho_{n,l}=\int\frac{d^3p}{(2\pi)^3p^0}\left(p^0\right)^n P_{2l}\left(\frac{p_z}{|\vec{p}|}\right)f(x^{\mu},p^{\mu})~,
    \label{intro}
\end{eqnarray}
    with $f$ as the single particle distribution function (with particle four-momenta $p^{\mu}$) and $P_{2l}$ is the Legendre polynomial with order $2l$. 
The equations of motion of $\rho_{n,l}$, which give the evolution dynamics for the moments derived from the kinetic theory, are observed to be highly coupled. In studies like \cite{Blaizot:2017ucy,deBrito:2024vhm} it has been shown that the evolution equations of the lower moments duly depend on that of higher order, making a physical truncation of this moment hierarchy inevitable. In previous studies, it has also been observed that with momentum-independent RTA,  only the indices $l$ of Eq.\eqref{intro} are coupled. However, the energy exponent $n$ does not mix and attributes individual equations of motion for each moment. In the current analysis, the situation becomes much more involved when the particle momentum dependence is introduced in the relaxation time of the collision term in kinetic equation. The moments $\rho_{n,l}$ now get coupled via the $n$ indices as well through the momentum dependence of MDRTA. Consequently,
the resulting chain of moment equations becomes much more interdependent. However, we argue that a momentum-dependent relaxation time approximation captures valuable dynamical information since (though via a simplistic model) it provides an account of the microscopic momentum transfer within the medium. The results corroborate our apprehension as we see that the inclusion of momentum dependence in the RTA formalism significantly impacts the behavior of the moments as the time evolution of their solution becomes sensitive to the energy dependence of the relaxation time under MDRTA.

The manuscript is organized as follows. In section II, the detailed formalism of the work has been derived in two
subsections. In subsection A, the framework has been set up by introducing the relativistic kinetic equation and 
different existing collision kernel used to solve it. Subsection B is dedicated to the derivation of the moment
evolution equations with MDRTA collision kernel pointing its dynamical difference over AW-RTA kernel. It also contains the computational complexities faced in solving the moment hierarchy and the possible solutions to it. Section III
contains the results and it depicts how the momentum dependent relaxation time actually impacts the solution of 
moments over the existing results. Finally, in section IV we conclude our study with summarizing our results and providing possible outlooks
for the current analysis.

%====================================================================================================
%====================================================================================================
\section{Formalism}
%----------------------------------------------------------------------------------------------------
\subsection{The relativistic transport equation}
In covariant kinetic theory, the microscopic behavior of a system of particles is characterized by the phase space distribution function $f(x^{\mu},p^{\mu})$, which, when multiplied with the appropriate phase space volume gives the probability of finding a particle at a point $(x^{\mu},p^{\mu})$ in phase space. The relativistic Boltzmann equation gives its evolution dynamics as the following,
\begin{equation}
    p^{\mu}\partial_\mu f_p(x^{\mu},p^{\mu}) = C[f,f]~,
    \label{kin}
\end{equation}
with $C[f,f]$ as the collision kernel that includes the system interactions. The Boltzmann equation is a non-linear integro-differential equation and therefore is both numerically and analytically challenging to solve. But to gain insight into the qualitative features of the solution we can approximate the collision kernel with model equations, the form of which is derived from the linearized operator,
\begin{equation}
    p^{\mu}\partial_\mu f(x^{\mu},p^{\mu}) ={\cal{L}}[\phi]~.
    \label{kin-lin}
\end{equation}
The relaxation time approximation (RTA) is one such very popular method that linearizes the collision kernel where the non-equilibrium distribution function restores its local equilibrium over a relaxation time scale (\(\tau_R\)) as the following,
\begin{equation}
    p^{\mu}\partial_\mu f(x^{\mu},p^{\mu}) = -\frac{p^{\mu}u_{\mu}}{\tau_R}f_p^{eq}~\phi~.
    \label{relax}
\end{equation}
Here, the non-equilibrium distribution function $f$ is decomposed in an equilibrium and an out-of-equilibrium deviation part as $f-f^{eq}=f^{eq}\phi$, and ${\cal{L}}[\phi]$ is the linearized collision operator of this deviation function $\phi$. The equilibrium distribution function $f^{eq}$ is given by,
\begin{align}
    f^{\text{eq}} = \frac{1}{e^{\left(p\cdot u - \mu\right)/T} + r}~,
    \label{dist-eq}
\end{align}
where $r$ takes values $\{-1,0,1\} $ corresponding to Bose-Einstein, Boltzmann, and Fermi-Dirac statistics, respectively. In the above definition, $u^{\mu}$ is a time like four vectors representing a local flow velocity, $T$ is the temperature and $\mu$ is the chemical potential of the system, respectively.

Now, in conventional RTA formalism given by Anderson and Witting (AW-RTA) \cite{Anderson:1974nyl}, the expression of $\tau_R$ in Eq.\eqref{relax} is independent of particle momenta $p^{\mu}$.
Hence, taking zeroth and first moment of \eqref{relax} readily gives the conservation of particle-4-flow $N^{\mu}$ and energy-momentum tensor tensor $T^{\mu\nu}$ with the given form of the ${\cal{L}}[\phi]$ (right hand side of Eq.\eqref{relax}) as long as the Landau matching condition of hydrodynamics is satisfied \cite{Romatschke:2009im,Denicol:2012cn}. Thus the AW-RTA Kernel is restricted to a specific choice of hydrodynamic frame. However, with a momentum dependent relaxation time approximation (MDRTA) (and with arbitrary hydrodynamic frames) the collision kernel given in Eq.\eqref{relax} faces serious issues of conservation violation and thus is unsuitable to study the non-equilibrium evolution of systems with MDRTA. The solution comes with the reformed collision kernels recently introduced and used in \cite{Rocha:2021zcw,Biswas:2022cla,Biswas:2022hiv,Teaney:2013gca}, that conserve the collision invariants irrespective of the momentum dependence in $\tau_R$ or the chosen hydrodynamic frame. The structural difference between the two referred collision kernels (under MDRTA) is merely because of the orthogonal basis in which the collision term is expanded. In the next subsection, when formulating the evolution dynamics of the moments, we will discuss this in more details. 
%-----------------------------------------------------------------------------------------------------
\subsection{Moment Evolution}
\subsubsection{Basic definitions and properties}
To study the evolution of bulk properties of the system, as mentioned in the introduction in Eq.\eqref{intro}, we define here the moments of the particle distribution function. But before that, we need to discuss the symmetries considered for the current analysis.
In this study we consider a simplified conformal system having rotational and translational symmetry in the transverse $x-y$ plane and boost invariance along $z$ axis. These symmetries can be conveniently manifested using the hyperbolic (Milne) coordinates given by the metric tensor, $g_{\mu\nu}=(1,-1,-1,-\tau^2)$ with proper time, $\tau=\sqrt{t^2 -z^2}$ and space-time rapidity, $\eta_s=\tanh^{-1}{z/t}$.
Under these constraints, the fluid flow profile reduces to $u^{\mu}  = (1,0,0,0)$. The distribution function then only has spacetime dependence via $\tau$, and momentum dependence through the transverse momentum $p_T$ and longitudinal momentum $p_z =p_{\eta}/\tau $.
Under these symmetry conditions, the moments of the particle distribution function take the following form
\cite{deBrito:2024vhm,Blaizot:2021cdv,Jaiswal:2022udf},
\begin{align}
    \rho_{n,l}(\tau) &= \int dP p^{n}P_{2l}(\cos{\theta}) f (\tau,p_T,p_\eta)~.
    \label{moment-def}
   \end{align}
Here $dP=\frac{d^3p}{(2\pi)^3p^0}$ is the phase space factor, $E=p^{0}=p$ is the single particle energy and $\cos{\theta} = p_z/p=p_z/|\vec{p}|=p_{\eta}/(\tau p^0)$.
In the moment expression of \eqref{moment-def}, the index $n$ measures energy scaling, where the index $l$ measures momentum anisotropy in the system. 

Next, we list here few of the properties of the moment expression \eqref{moment-def}.
The moment corresponding to equilibrium distribution function $\rho_{n,l}^{eq}$ has the following form,
\begin{align}
    \rho_{n,l}^{eq}(\tau) &= \int dP p^{n}P_{2l}(\cos{\theta}) f^{eq}(\tau,p_T,p_\eta)~,
    \label{momeq-def}
\end{align}
with the expression of $f_p^{eq}$ taken from \eqref{dist-eq} (we have used the Maxwell-Juttner distribution). Noticing that $f_p^{eq}$ is independent of $\theta$ (only function of $p$), we see that by the virtue of orthogonality property of Legendre polynomial $\int_{-1}^{1}P_m(x)P_n(x)dx\sim \delta_{mn}$, we have, 
\begin{equation}
     \rho_{n,l}^{eq}(T,\mu)  = e^{\mu/T}\frac{T^{n + 2}}{2\pi^2}\Gamma(n+2)\delta_{l0}~,
     \label{moment-def1}
\end{equation}
where $\Gamma(n)=(n-1)!$ is the gamma function and $\delta_{ij}$ is the Kronecker delta function.
From Eq.\eqref{moment-def1} we can see that because of the Legendre polynomial properties, the equilibrium moments vanish unless $l=0$.
It is to be noted here that the moments $\rho_{1,0}^{eq}$  and $\rho_{2,0}^{eq}$ correspond to the
equilibrium number density and equilibrium energy density, respectively.

The covariant kinetic equation Eq.\eqref{kin-lin} can be further simplified as,
\begin{align}
\partial_{t}f_p+\vec{v_p}\cdot\vec{\nabla}f_p=\frac{1}{p^0}{\cal{L}}[\phi_p]~,
\end{align}
with $\vec{v}_p=\vec{p}/p$. Under Bjorken symmetry it takes the following form \cite{Baym:1984np},
\begin{align}
\left[\frac{\partial}{\partial\tau}-\frac{p_z}{\tau}\frac{\partial}{\partial p_z}\right]f_p=-\frac{1}{p}{\cal{L}}[\phi_p]~.
\label{kin-Bjorken}
\end{align}
Multiplying Eq.\eqref{kin-Bjorken} with $p^nP_{2l}({\text{cos}}\theta)$ and integrating over $dP$,
we finally obtain the equation of motion governing the moment evolutions,
\begin{align}
\frac{\partial}{\partial\tau}\rho_{n,l} ~ + ~
&\frac{1}{\tau} \left[{\cal{P}}(n,l)\rho_{n,l-1}+{\cal{Q}}(n,l)\rho_{n,l}+{\cal{R}}(n,l)\rho_{n,l+1}\right]\nonumber\\
&=\int dP \left(p\right)^{n-1} P_{2l} (\text{cos}\theta) {\cal{L}}[\phi_p]~,
\label{moment2}
\end{align}
with,
\begin{align}
    {\cal{P}}(n,l)=&2l\frac{(2l-1)}{(4l-1)}\frac{(n+2l)}{(4l+1)}~,\nonumber\\
    {\cal{Q}}(n,l)=&\frac{2}{3} + \frac{n(8l^2+4l-1)}{(4l-1)(4l+3)}+\frac{2l(2l+1)}{3(4l-1)(4l+3)}~,\nonumber\\
    {\cal{R}}(n,l)=&(n-2l-1)\frac{(2l+1)}{(4l+1)}\frac{(2l+2)}{(4l+3)}~.
   \end{align}
   In deriving Eq.\eqref{moment2}, apart from the orthogonality property, the used the recurrence relations and the derivative properties of Legendre polynomial are listed below,
   \begin{align}
       &(4l+1)xP_{2l}(x)=(2l+1)P_{2l+1}(x)+2l~P_{2l-1}(x)~,\\
       &\partial P_{2l}(x)/\partial x=2l\left\{P_{2l-1}(x)-xP_{2l}(x)\right\}/(1-x^2)~.
   \end{align}
For the conventional momentum independent AW-RTA, where the collision kernel is given by Eq.\eqref{relax}, the right hand side of moment equation \eqref{moment2} can be trivially simplified
to,
\begin{align}
    \frac{\partial}{\partial\tau}\rho_{n,l} ~ + ~
&\frac{1}{\tau} \left[{\cal{P}}\rho_{n,l-1}+{\cal{Q}}\rho_{n,l}+{\cal{R}}\rho_{n,l+1}\right]\nonumber\\
&=-\frac{1}{\tau_R}\left[\rho_{n,l}-\rho_{n,l}^{eq}\right]~,
\label{moment3}
\end{align}
which is result derived in \cite{deBrito:2024vhm}. For $n=2$, the results agree with that of \cite{Blaizot:2017ucy} as well.
%---------------------------------------------------------------------------------------------------
\subsubsection{Collision kernel with the momentum-dependent relaxation time approximation}

The problem of a system with a momentum-dependent relaxation time has been investigated earlier using hydrodynamic equations \cite{Teaney:2013gca,Kurkela:2017xis,Rocha:2021zcw,Mitra:2020gdk,Dash:2021ibx,Dash:2023ppc,Singh:2024leo,Bhadury:2024ckc,KumarSingh:2025kml,Mukherjee:2025dqp}. We investigate this from the perspective of kinetic theory, which gives insight into how such a system approaches equilibrium. Accordingly, while treating the collision term in the moment equation \eqref{moment2}, we use the novel relaxation time collision operator \cite{Rocha:2021zcw} that conserves both particle number and energy-momentum irrespective of the momentum-dependence considered in $\tau_R(p)$ and the hydrodynamic frame chosen,
\begin{align}
    {\cal{L}}[\phi_p]=&-\frac{p}{\tau_R}f^{eq}\Bigg[\phi_p-P_0\frac{{\langle\frac{p}{\tau_R}P_0\phi_p\rangle}_{eq}}{{\langle\frac{p}{\tau_R}P_0P_0\rangle}_{eq}}\nonumber\\
    &-P_1\frac{{\langle\frac{p}{\tau_R}P_1\phi_p\rangle}_{eq}}{{\langle\frac{p}{\tau_R}P_1P_1\rangle}_{eq}}
    -p^{\langle\mu\rangle}\frac{{\langle\frac{p}{\tau_R}p_{\langle\mu\rangle}\phi_p\rangle}_{eq}}{\frac{1}{3}{\langle\frac{p}{\tau_R}p_{\langle\nu\rangle}p^{\langle\nu\rangle}\rangle}_{eq}}\Bigg] ~,
    \label{MDRTA-coll}
\end{align}
with $P_0=1$ and $P_1=1-p{\langle\frac{p}{\tau_R}\rangle}_{eq}/{\langle\frac{p^2}{\tau_R}\rangle}_{eq}$. The notation ${\langle\cdots\rangle}_{eq}$ reads, ${\langle\cdots\rangle}_{eq}=\int dP f_p^{eq} (\cdots)$.
It should be mentioned here that a slightly different looking collision kernel is introduced in \cite{Biswas:2022cla}
based on the chosen momentum basis in which $\phi_p$ is expanded. However, it can be trivially seen that a momentum  
rearrangement proves their equivalence. Here, because of the neatness in the expression of collision term we proceed 
with ${\cal{L}}[\phi_p]$ given in \eqref{MDRTA-coll}.

With the given prescription, we denote the right-hand side of Eq.\eqref{moment2} (moment of the linearized collision term) as $C_{n,l,\Lambda}$ such that the moment evolution equation becomes,
\begin{eqnarray}
    \frac{\partial\rho_{n,l}}{\partial\tau} = &&-  \frac{1}{\tau}\bigg[\mathcal{P}(n,l)\rho_{n,l-1} + \mathcal{Q}(n,l) \rho_{n,l} + \mathcal{R}(n,l)\rho_{n,l+1} \bigg]\nonumber\\
    &&~~+~~  C_{n,l,\Lambda}~.
    \label{Eq:MomEvol}
\end{eqnarray}

To proceed further, we decompose the momentum-dependent relaxation time $\tau_R(p)$ into a momentum-independent part $\tau_R$ (precisely a function of temperature $T$) and a part purely a function of single particle energy $p$ with the exponent $\Lambda$ being a positive number \cite{Dusling:2009df,Mitra:2020gdk},
\begin{equation}
\tau_R(p) = \tau^{0}_{R}(T)p^{\Lambda}~.
\label{MDRTA}
\end{equation}
The momentum independent part can be simply calculated using the shear viscosity ($\eta$) over entropy density ($s$) ratio of the system as the following \cite{Mitra:2020gdk},
\begin{equation}
    \tau^{0}_{R}(T) =\lrb{\frac{\eta}{s}}\frac{5!}{\Gamma(5+\Lambda)}\frac{1}{T^{1+\Lambda}}~.
    \end{equation}
From now on, whenever the notation $\tau_R$ has been mentioned either in the text or in the figures of the result section, it indicates this thermal part and must not be confused with the total relaxation time $\tau_R$.

Using expression \eqref{MDRTA}, the moment of the collision term $C_{n,l,\Lambda}$ becomes (Appendix \eqref{Apx:ColKer}),
\begin{align}
    C_{n,l,\Lambda}=&-\frac{1}{\tau^{0}_{R}} \Bigg[\left\{\rho_{n-\Lambda,l}-\delta_{l0}~\rho^{eq}_{n-\Lambda,l}\right\}
    -A~\delta_{l0}~\rho^{eq}_{n-\Lambda,0}\nonumber\\
    &-B~\delta_{l0}~\left\{\rho^{eq}_{n-\Lambda,0}-\frac{\rho^{eq}_{1-\Lambda,0}}{\rho^{eq}_{2-\Lambda,0}}\rho^{eq}_{n-\Lambda+1,0}\right\}    \Bigg]~,
    \label{coll2}
\end{align}
with,
\begin{align}
    A=&\frac{\rho_{1-\Lambda,0}-\rho^{eq}_{1-\Lambda,0}}{\rho^{eq}_{1-\Lambda,0}}~,\\
    B=&\frac{\rho_{1-\Lambda,0}-\frac{\rho^{eq}_{1-\Lambda,0}}{\rho^{eq}_{2-\Lambda,0}}\rho_{2-\Lambda,0}}
    {\rho^{eq}_{3-\Lambda,0}\left(\frac{\rho^{eq}_{1-\Lambda,0}}{\rho^{eq}_{2-\Lambda,0}}\right)^2-\rho^{eq}_{1-\Lambda,0}}~.
\end{align}
With the values of with $l=0$ and $n=1,2$ it can be observed that $C_{n,l,\Lambda}$ readily gives zero, preserving the collisional invariant property (conservation laws) independent of $\Lambda$. The resulting equations give the well-known conservation equations for particle number $({\rm n})$ and energy density $(\epsilon)$ for a system as follows,
\begin{align}
&\frac{\partial {\rm n}}{\partial\tau}+\frac{{\rm n}}{\tau}=0~,\\
&\frac{\partial\epsilon}{\partial\tau}+\frac{4}{3}\frac{\epsilon}{\tau}=-\frac{2}{3}\frac{P_L-P_T}{\tau}~,
\end{align}
where we identify the moments as, $n=\rho_{1,0}~,~ \epsilon=\rho_{2,0}$ and $\rho_{2,1}=P_L-P_T=$ pressure anisotropy.

After some tedious algebra, the collision moment $C_{n,l,\Lambda}$ can be written in the following simplified form(Appendix \eqref{Apx:ColKer}),
\begin{align}
    C_{n,l,\Lambda}= &-\frac{1}{\tau_R^{0}} \Big[\rho_{n-\Lambda,l}\nonumber\\
    &- \delta_{l0}\Big\{ T^{n-1}K(n,1,\Lambda)[1 - C(n,\Lambda)]\rho_{1-\Lambda,l} \nonumber\\
    &~~~~~~+ T^{n-2}K(n,2,\Lambda)C(n,\Lambda)\rho_{2-\Lambda,l}\Big\}\Big]~,
     \label{coll3}
    \end{align}
with the following two functions introduced,
\begin{align}
    & K(n,m,\Lambda)= \frac{\Gamma(n-\Lambda +2)}{\Gamma(m-\Lambda +2)} ~,\\
    & C(n,\Lambda)= \frac{1 - K(1,n,\Lambda)K(n+1,2,\Lambda)}{1 - K(1,2,\Lambda)K(3,2,\Lambda) }~,
\end{align}
such that $C(1,\Lambda) = 0$,  $C(2,\Lambda) = 1$ and $K(n,n,\Lambda) = 1$. This makes energy and number conservation explicit as $C_{n,l,\Lambda}$ is zero for $n = 1,2$ and $l=0$. Here, the temperature $T$ and chemical potential $\mu$ is defined via the Landau conditons $\epsilon = \epsilon_{\text{eq}}$ and ${\rm n} = {\rm n}_{\text{eq}}$,
\begin{align}
    {\rm n} &= e^{\mu/T}\frac{T^{3}}{2\pi^2}\Gamma(3)\,,\\
    \epsilon &= e^{\mu/T}\frac{T^{4}}{2\pi^2}\Gamma(4)\,.
\end{align}

When $\Lambda$ is zero,  the moments are coupled to each other only through the $l$ indices while for any non-zero $\Lambda$, each $(n,l)$ moment is coupled to the corresponding $(n-\Lambda,l)$ moment via the collision term $C_{n,l,\Lambda}$ (see Eq.\eqref{coll3}). This creates an infinite ladder of coupling towards the lower $n$ moments.

In the following, we are writing the derived moment evolution equations of the current analysis using a momentum-dependent relaxation time collision kernel,
\begin{align}
&\text{For~~}~ l\neq0 ~,\nonumber\\
&\frac{\partial}{\partial\tau}\rho_{n,l}+\frac{1}{\tau}\left[{\cal{P}}\rho_{n,l-1}+{\cal{Q}}\rho_{n,l}+{\cal{R}}\rho_{n,l+1}\right]=-\frac{1}{\tau^{0}_R}\rho_{n-\Lambda,l}~,
\label{moment-final1}\\
&\text{For~~}~ l=0 ~,\nonumber\\
&\frac{\partial}{\partial\tau}\rho_{n,0}+\frac{1}{\tau}\left[\frac{2}{3}(n-1)\rho_{n,1}+\frac{1}{3}(n+2)\rho_{n,0}\right]=\nonumber\\
&~~-\frac{1}{\tau^{0}_R}\Big[\rho_{n-\Lambda,0}-T^{n-1}K(n,1,\Lambda)\left\{1-C(n,\Lambda)\right\}\rho_{1-\Lambda,0}\nonumber\\
&~~~~~~~~~~~~~~~~~~~-T^{n-2}K(n,2,\Lambda)C(n,\Lambda)\rho_{2-\Lambda,0}\Big]~.
\label{moment-final2}
\end{align}
Eq.\eqref{moment-final1} is structurally not very different from the momentum-independent case \eqref{moment3} apart from the $\Lambda$ factor inclusion in the $n$ index of the moment to the source term at the right hand side (however, that change the dynamics anyway). But it is the $l=0$ moment in \eqref{moment-final2} that bears the effect of MDRTA
the most via the $C(n,\Lambda)$ and $K(m,n,\Lambda)$ functions. It again carries its effect recursively through the couplings of the moment hierarchy as mentioned earlier.

\section{Computational Details}

\subsection{Initial Conditions}

We consider a conformal system with particle mass $m=0$ at an initial temperature $T(\tau_0) = 1$ GeV and initial chemical potential $\mu(\tau_0) = 0$. We initialize the moments $\rho_{n,l}(\tau_0)$ to their equilibrium values $\rho^{eq}_{n,l}$. Unlike momentum-independent relaxation time, systems with larger viscosity cannot be probed by considering smaller values of initial rescaled time. In this study, we choose two values of the initial viscosity to entropy ratio $\eta_0/s_0 = 0.2,0.02$. To make a comparison with the previous works \cite{deBrito:2024vhm}, we then choose the initial time $\tau_0$ such that the scaled time is $\tau_0/\tau_R(\tau_0) \in \{ 0.10,~ 1.0\}$ so that the initial parameter set matches the case for $\Lambda = 0$. We then solve the coupled equations in \eqref{Eq:MomEvol} by truncating the moment equations at  $(n_{\rm min},n_{\rm max}) = (0,5)$,  and $(l_{\text{min}},l_{\text{max}})  =(0,100)$ using the RK4 algorithm.

\subsection{ Truncation: Free Streaming}
\begin{figure}[!h]
    \centering
    \includegraphics[width=0.8\linewidth]{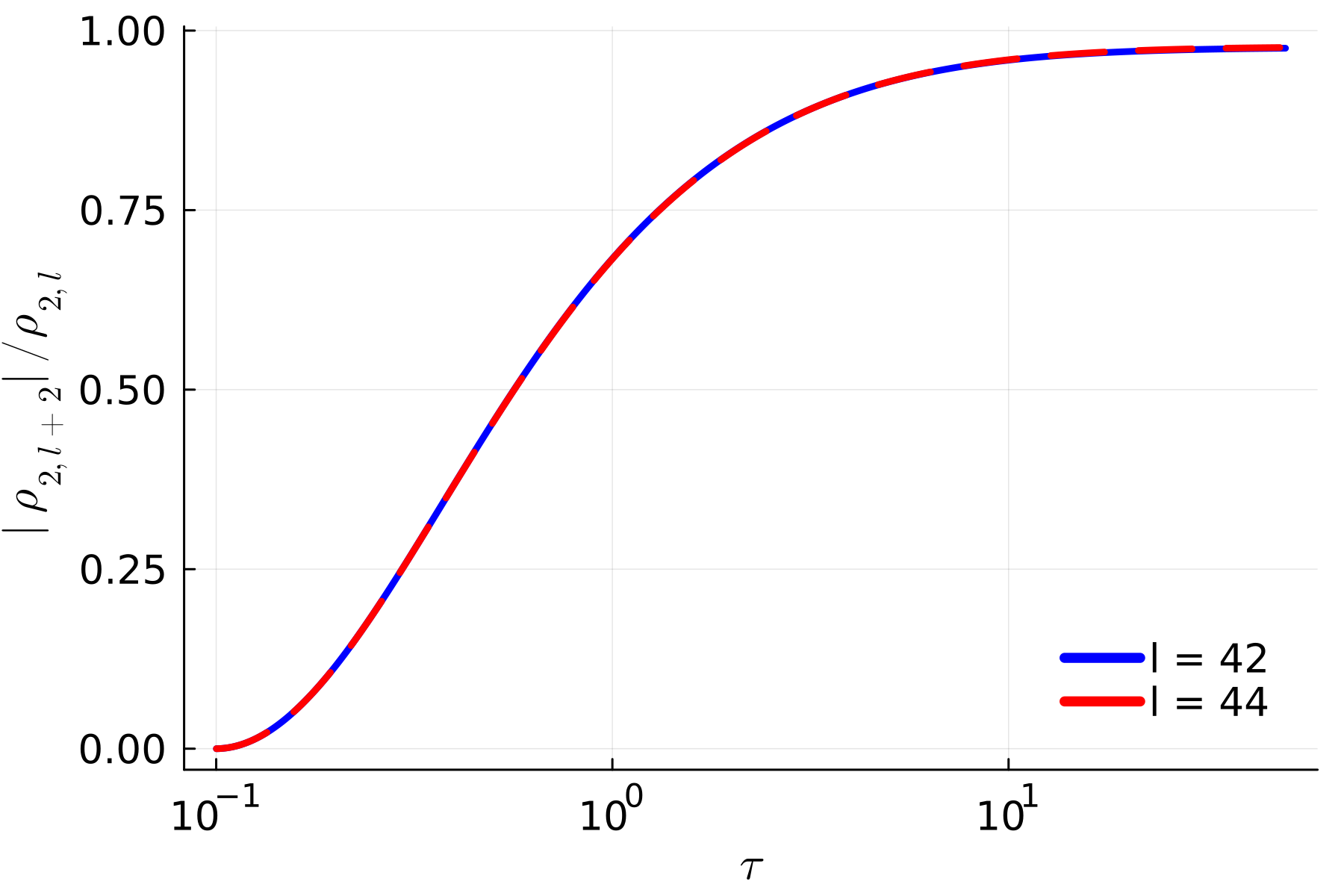}
    \caption{The ratio of absolute values of alternate moments at late time for $n= 2$ for a free streaming system. The ratio  saturates near 1. }
    \label{fig:ratio}
\end{figure}

The moment evolution equation \eqref{Eq:MomEvol} is a set of infinitely coupled non-linear differential equations. To compute the moments, for practical purposes we need to truncate these equations at some finite order.  In a previous work by de Brito et.al, \cite{deBrito:2024vhm} the $l$ moments were truncated by setting $\rho_{n,l} = 0$ for some $l \geq l_{max}$, note that due to momentum independent relaxation time there was no $n$ coupling. This assumes that the contribution of the $l$ moments beyond $l_{max}$ can be ignored compared to the lower $l$ moments. But this type of truncation scheme no longer holds true for later times, when the $l$ moments decay and become comparable in value with respect to each other. The situation is elaborated below by analyzing different terms of Eq.\eqref{Eq:MomEvol}. The free streaming evolution equation is given by,
\begin{align}
    \frac{\partial}{\partial\tau}\rho_{n,l}=& -\frac{1}{\tau}{\cal{Q}}(n,l)~\rho_{n,l}\nonumber\\
    &- \frac{1}{\tau}\left[{\cal{P}}(n,l)~\rho_{n,l-1}+{\cal{R}}(n,l)~\rho_{n,l+1}\right]\,.
    \label{compare}
\end{align}
In the first term on the right-hand side of \eqref{compare}, ${\cal{Q}} > 0$, indicating that this is a decay term. The terms ${\cal{P}}$ and  ${\cal{R}}$ have opposite signs while $\rho_{n,l-1}$ and $\rho_{n,l+1}$ share the same sign.  On the other hand \(\rho_{n,l}\) alternates sign with \(l\), meaning that $\rho_{n,l}$ has opposite sign of $\rho_{n,l-1}$ and $\rho_{n,l+1}$. Initially, all the moments except \(l=0\) are zero. It is observed that for the given initial condition, $\rho_{n,l+1} \ll \rho_{n,l-1}$ during the early time and the ${\cal{P}}(n,l)~\rho_{n,l-1}$ (see Fig.\eqref{fig:ratio}) term is non-zero and contributes to a coupling between \(l\) modes. Noting that for large values of $l$, the coefficients in the evolution equation is,
\begin{align}
    P(n,l\gg1) &\sim \frac{l}{2} + \frac{(n+1)}{4}\\
    Q(n,l\gg1) &\sim \frac{(2n+3)}{4}\\
    R(n,l\gg1) &\sim  -\frac{l}{2} + \frac{(n-2)}{4}\,.   
\end{align}
 At late time we have $\rho_{n,l+1} \sim \rho_{n,l-1}$, and eventually the last two terms on the right-hand side of Eq.\eqref{compare} partially cancel each other and for a large enough $\tau$. Therefore a better truncation for the free streaming evolution can be obtained by setting $\rho_{n,l_{\text{max}}-1} = \rho_{n,l_{\text{max}}+1} $.

\subsection{Truncation: MDRTA}
Now, let us consider the case with a non-zero collision kernel. The truncation for the momentum-dependent is non-trivial as there is coupling in both $l$ and $n$ indices (see \eqref{coll3}). A non-zero \(\Lambda\) fundamentally changes the type of the coupled differential equation. To understand this, let us rewrite the set of coupled differential equations in the following matrix representation 
 \begin{equation}\label{Eq:MomMat}
     \dv{\vec{\rho}}{\tau}= -\frac{1}{\tau} \mathbf{F} \vec{\rho} -\frac{1}{\tau^{0}_R}{\mathbf{C}}(\Lambda)\vec{\rho}\,,
 \end{equation}
where $\vec{\rho}$ is a vector consisting of the infinite (or finite for a truncated system) set of $(n,l)$ moments and $ \mathbf{F}$ and $\mathbf{C}(\Lambda)$ are matrices. The vector $\vec{\rho}$ is a collection of moments $\rho_{n,l}$, $n_{min}<n< n_{max}$, $0<l< l_{max}$, which are given some arbitrary ordering  and written in a vector form $\vec{\rho} = (\rho_{0,0},\rho_{0,1}\dots,\rho_{1,0},\rho_{1,1}\dots)$. The matrices $\mathbf{F}$ and $\mathbf{C}(\Lambda)$ couples these moments. Specifically $\mathbf{F}$ couples the $l$ moments for a fixed $n$, while $\mathbf{C}(\Lambda)$ couples the $n$ moments(See Appendix \eqref{Ap:Vec} ).

\begin{figure}[!h]
    \centering
    \includegraphics[width=0.8\linewidth]{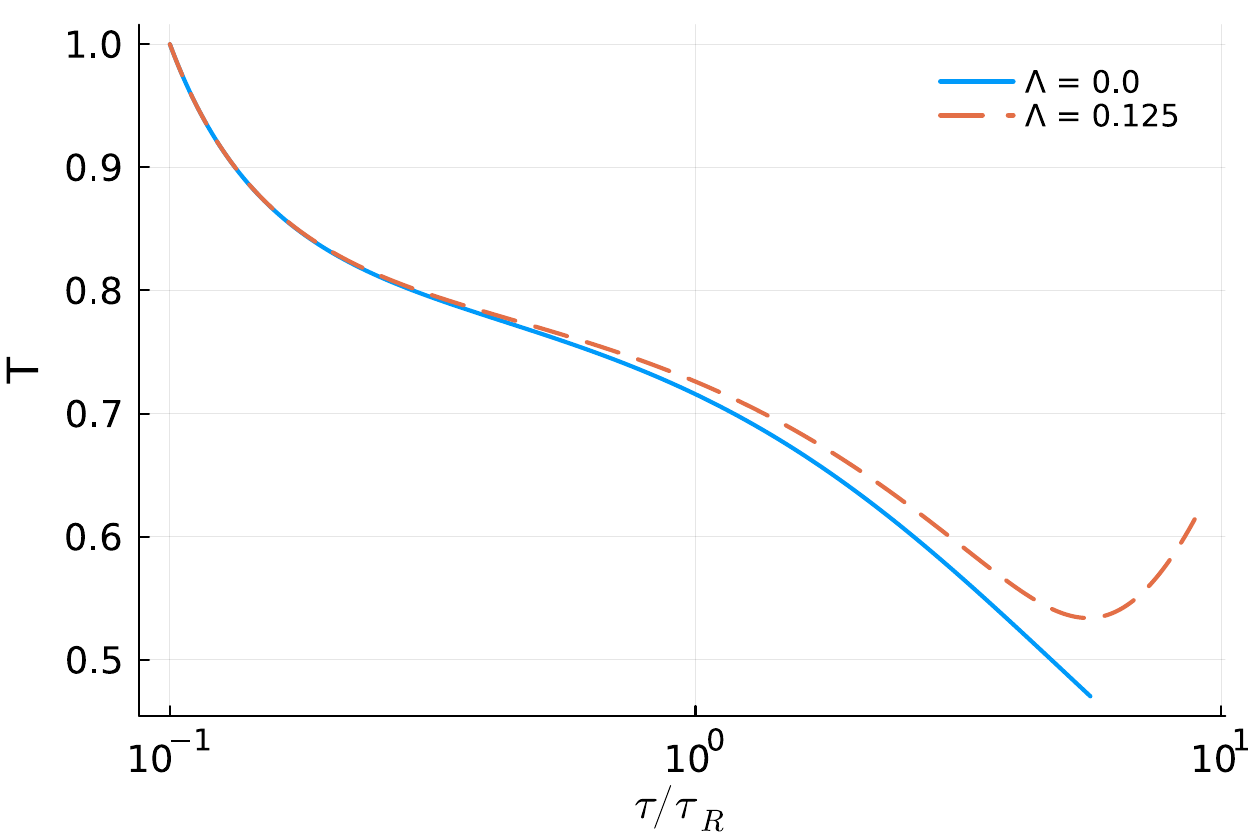}
    \caption{The evolution of temperature for $\Lambda = 0$ and $\Lambda = 0.125$. We see a growth in temperature when $\tau/\tau_R > 1$ .}
    \label{fig:TempDiv}
\end{figure}

The matrix $\mathbf{F} $ gives free streaming dynamics and is dominant when $\tau/\tau_R \ll 1$.  The second matrix $\mathbf{C}(\Lambda)$ gives the collision dynamics. The presence of $1/\tau_R \sim T^{1+\Lambda} $, makes the above differential equation nonlinear. However, the local (in time) dynamical behavior of the differential equation can be inferred from the eigenvalues of $\mathbf{C}(\Lambda)$. For late times, i.e $\tau/\tau_R > 1$, we can assume the dynamics to be controlled by $\mathbf{C}(\Lambda)$. When the collision term is dominant, we can ignore the free streaming term $\mathbf{F}/\tau $ and  write a formal solution to Eq.\eqref{Eq:MomMat} as,
 \begin{align}\label{Eq:FormSol}
     \vec{\rho}_n(\tau) \sim \exp\lrrb{{-\int d\tau\frac{1}{\tau^{0}_R}\mathbf{C}(\Lambda)} }\vec{\rho}_n(\tau_0)~,  && \tau/\tau_R \gg 1.
 \end{align}
For $\Lambda=0$, the matrix is diagonal ($\mathbf{C}(\Lambda) \sim I$) and eigenvalues are positive and give rise to a pure decay.  However, for non-zero \(\Lambda\), $\mathbf{C}(\Lambda)$ (see Eq.\eqref{eq:Ctrunc} for n = 4 truncation) is off-diagonal and is similar to the left shift operator,
\begin{align}\label{eq:Ctrunc}
    \mathbf{C}(\Lambda) \sim\begin{bmatrix}
        0 && 0 && 0&& 0\\
        c_{n-4\Lambda} && 0 && 0&& 0\\
        0 && c_{n-3\Lambda} && 0&& 0\\
        0 && 0 && c_{n-2\Lambda}&& 0
    \end{bmatrix}\,.
\end{align}
For finite truncations, the spectrum of the operator $\mathbf{C}(\Lambda)$, is no longer positive definite, and a late-time decay is not guaranteed. In Fig.\eqref{fig:TempDiv}, the evolution of temperature for two different values of $\Lambda \in \{0,0.125\}$ are given. We see that the temperature evolution diverges from the expected late-time decay behavior for \(\Lambda =0.125\). 

As the evolution of the system at late times is governed by the eigenvalues of the matrix $\mathbf{C}(\Lambda)$, we can understand the divergent behavior by analyzing its spectrum. As $\textbf{C}(\Lambda)$ is lower triangular, all its eigenvalues are zero and, therefore, $\mathbf{C}^{N}(\Lambda) = 0$ (nilpotent), where $N$ is the dimension of the matrix. If $\mathbf{C}(\Lambda)$ is independent of time, the solution to Eq\eqref{Eq:FormSol} is,
\begin{equation}
    \Vec{\rho}(\tau) = \left(\sum_{k=0}^{N-1}\mathbf{C}(\Lambda)^{k}\tau^{k} \right)\Vec{\rho} (\tau=\tau_0)
\end{equation}
and therefore has polynomial growth instead of the expected exponential decay. This is the origin of the temperature growth seen in Fig.\eqref{fig:TempDiv}. 

\subsubsection{Moment Closure}
We know that the linearized collision kernel $\textbf{L}$ for the Boltzmann equation,
\begin{equation}
    p^{\mu}\partial_{\mu}f = - \textbf{L}(f-f^{eq})\,,
\end{equation}
is positive semi-definite \cite{cercignani2013mathematical} and the zero eigenvalues of the linearized operator correspond to the collision invariants or the conserved quantities. This effectively enshrines the approach to equilibrium and entropy generation in the system.  We expect this property to be preserved for any moment scheme representing the Boltzmann equation.

The divergence of temperature in Fig.\eqref{fig:TempDiv} is due to improper truncation and closure of the moments. A proper truncation scheme should capture the salient features of the infinite hierarchy of moments. To recover the positive definiteness of the collision kernel, we use the Grad moment closure procedure. Grad's procedure involves approximating the distribution function using a series expansion in momentum,
\begin{equation}\label{Eq:GradExp}
    f(\tau,p,\cos{\theta}) = f^{eq}\lrb{ \sum_{ m=0}^{N}\sum_{l=0}^{L} A_{m,l}(\tau)p^{m}P_{2k}(\cos{\theta})}\,,
\end{equation}
where $A_{m,l}$ are expansion coefficients and the series is truncated at $N$ and $L$.  We note that the Grad expansion in \eqref{Eq:GradExp} is not in general convergent\cite{deBrito:2024qow}. However, the series is asymptotic and can give accurate results for lower-order $p$ moments and when the system is near equilibrium. In our work we used $N=5$. The results were not found to be sensitive to the truncation order.

If we know the coefficients $A_{m,l}$, we can find any moment by integrating over the distribution function. Using the definition of the moments in \eqref{moment-def} we get the set of relations between the integer moments $\rho_{n,l}$ and $A_{m,l}$ via the vector equations (Appendix \eqref{Apx:MomCls}),
\begin{align}
     \Vec{\rho}_{l} &= \textbf{H}(T,0)\Vec{\text{A}}_{l} \,,
\end{align}
where $\textbf{H}(T,\Lambda)$ is an $N\times N$ matrix, $\Vec{\text{A}}_{l}$ and $\Vec{\rho}_{l}$ are column vectors of moments $\rho_{n,l}$ and coefficients $A_{m,l}$  constructed by fixing $l$ and varying $n$ (Detailed definitions are given in Appendix\eqref{Apx:MomCls}). This set of linear equations can be inverted to get $\text{A}_{m,l}$
\begin{align}
    \Vec{\text{A}}_{l} &= \textbf{H}^{-1}(T,0)\Vec{\rho}_{l}\,.
\end{align}
So knowing any $N\times L$ moments completely determines the coefficients $\text{A}_{m,l}$. Using this we approximate the fractional moments $\rho_{n-\Lambda}$ as
\begin{align}
    \Vec{\rho}_{n-\Lambda} &= T^{\Lambda}\textbf{H}(T,-\Lambda)\textbf{H}^{-1}(T,0)\Vec{\rho}_{n}\,,
\end{align}
which expresses the fractional moments as a linear combination of the integer moments.
The vector form of the moment equations now reads,
 \begin{equation}
     \frac{d}{d\tau}\vec{\rho}_{l}= -\frac{1}{\tau} \mathbf{F} \vec{\rho}_{l} -\frac{1}{\tau^{0}_R}\textbf{M}(T,\Lambda)\vec{\rho}_{l}\,.
 \end{equation}
 where $\textbf{M}(T,\Lambda) = T^{\Lambda}\textbf{H}(T,-\Lambda)\textbf{H}(T,0)$. One can verify numerically that $\textbf{M}$ has positive definite eigenvalues for an appropriate truncation in $N$. An analytical proof of the positive definiteness of $\textbf{M}(T,\Lambda)$ is given in \eqref{Apx:MomCls}.

\section{Results and discussions}
\begin{figure}[!h]
 \begin{subfigure}[!h]{1\textwidth}
    \centering
    \includegraphics[width=0.40\linewidth]{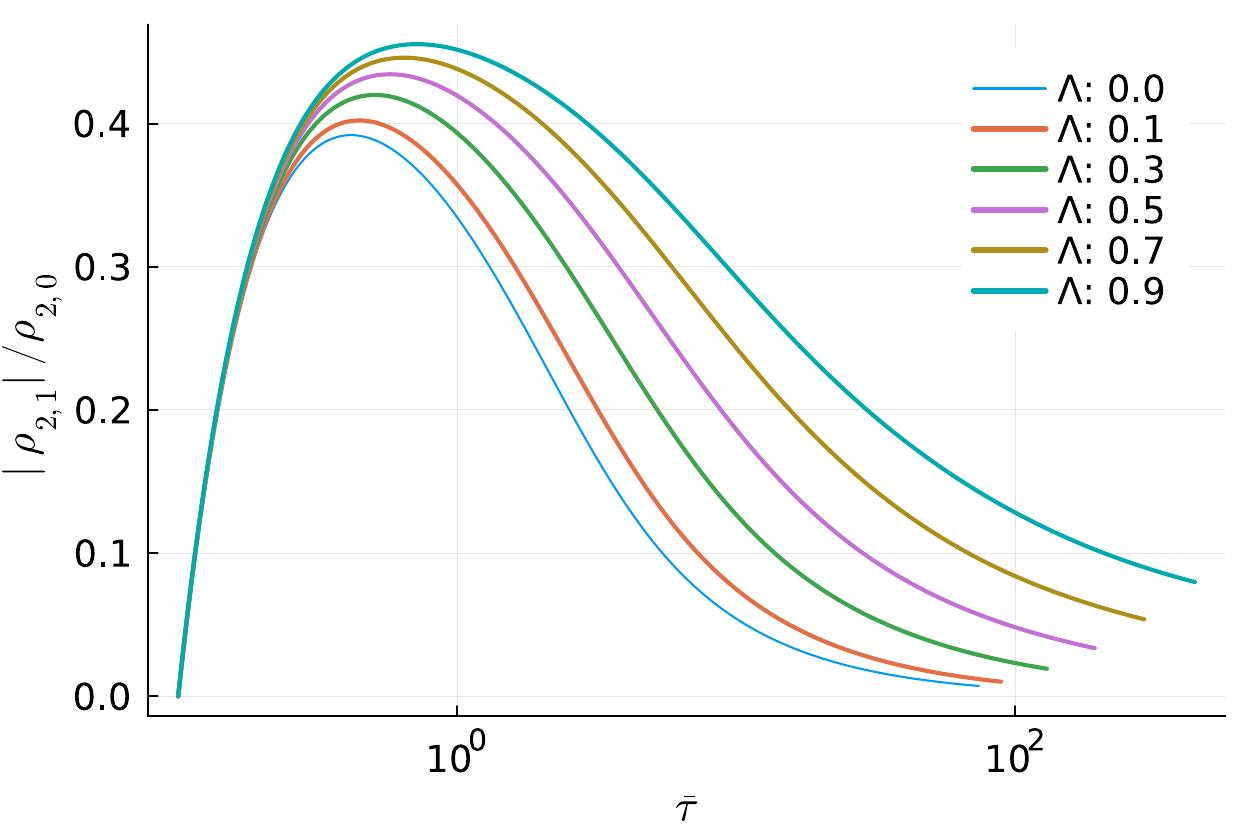} 
    \includegraphics[width=0.40\linewidth]{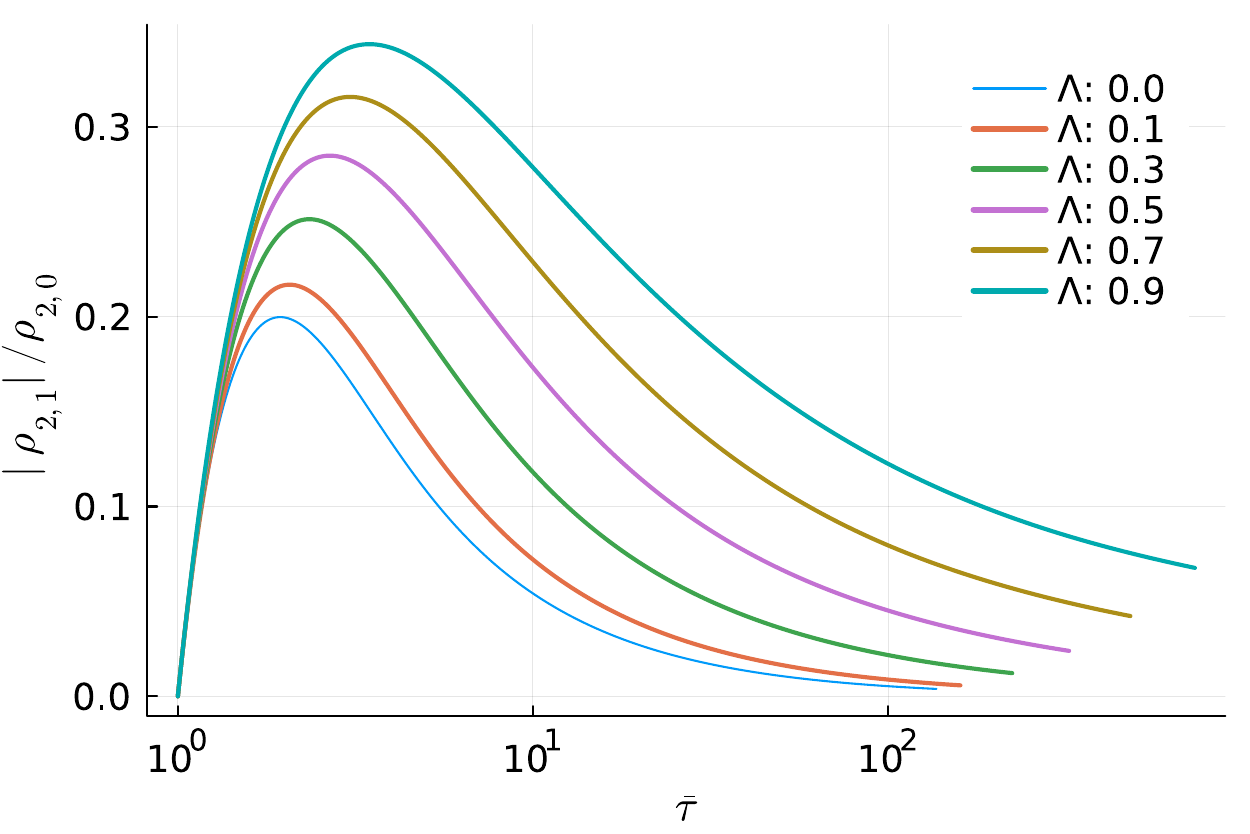} 
    \caption{Evolution of $\rho_{2,1}/\rho_{2,0}$ for various $\Lambda$ values and $\eta/s = 0.2,0.02$.}
    \label{fig:Rho21}
     \end{subfigure}
    \begin{subfigure}[!h]{1\textwidth}
    \centering 
    \includegraphics[width=0.40\linewidth]{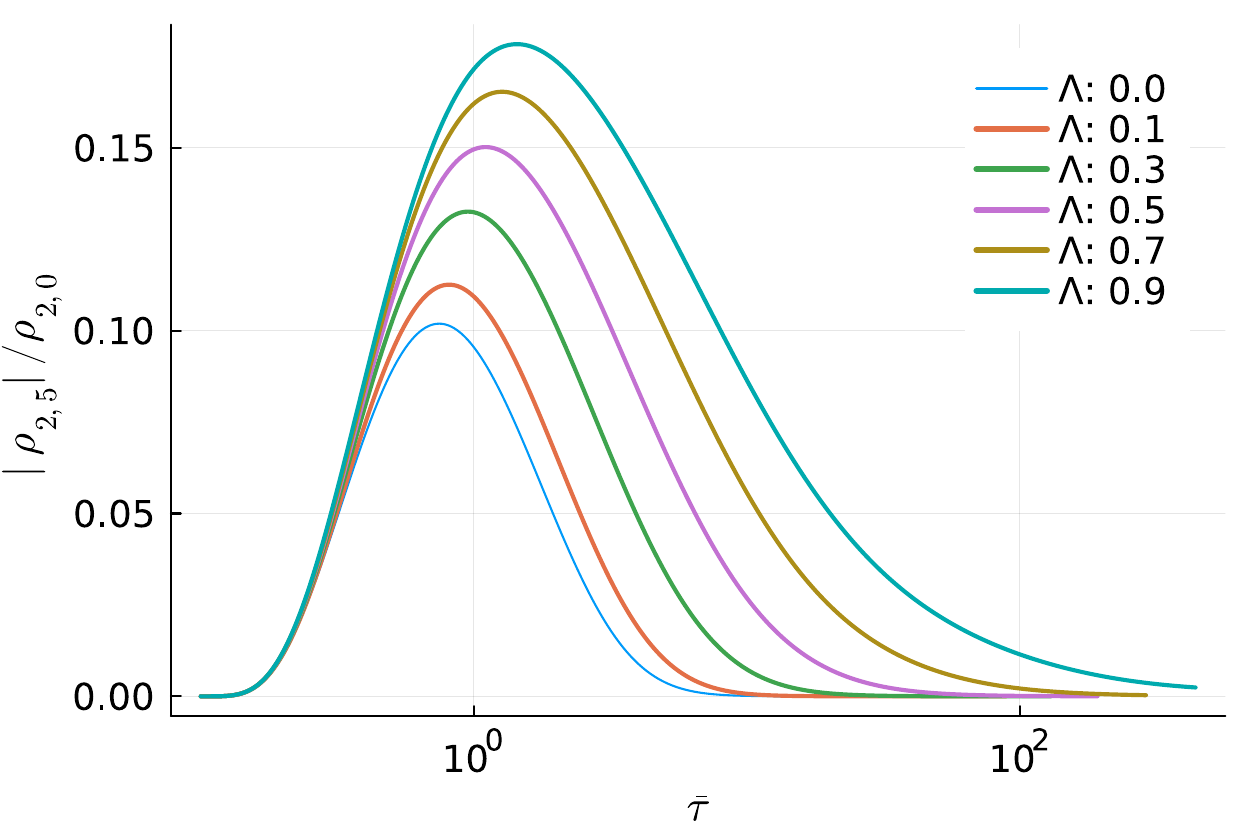} 
    \includegraphics[width=0.40\linewidth]{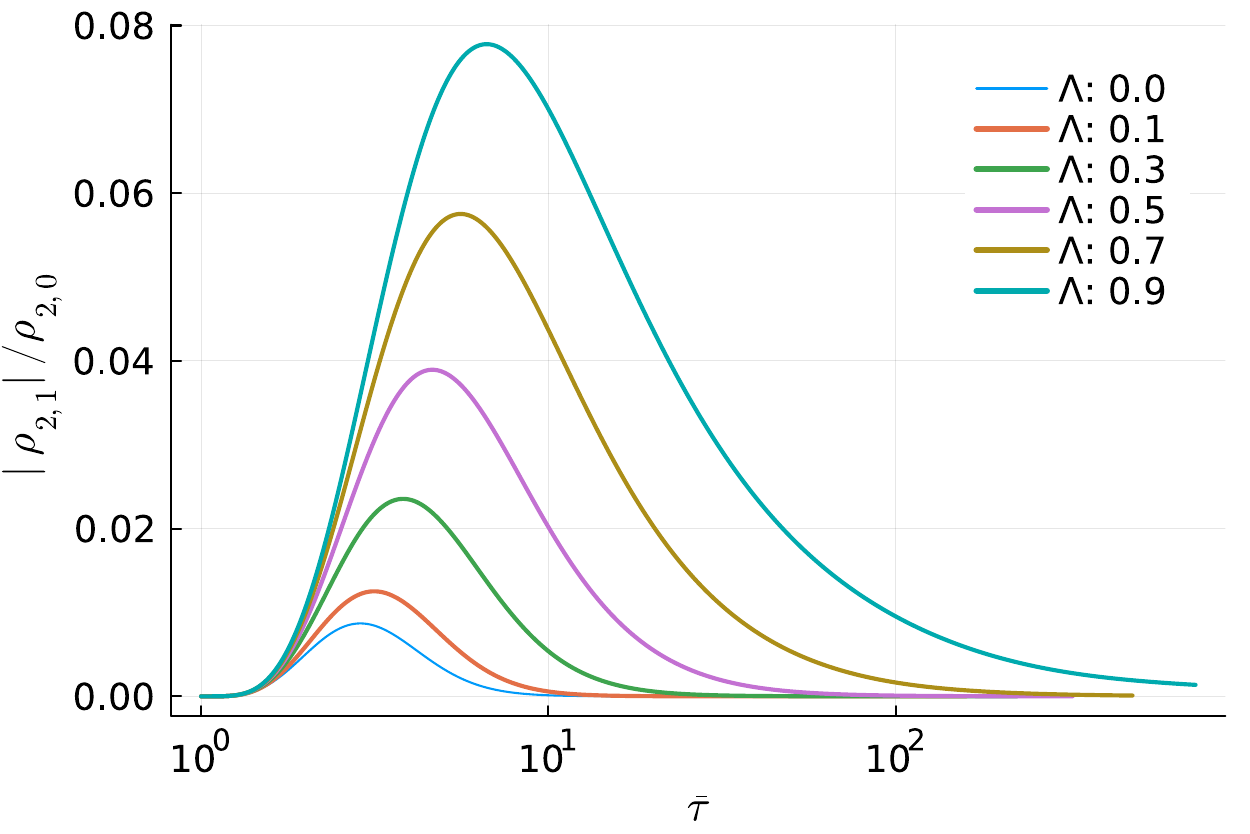}
    \caption{Evolution of $\rho_{2,5}/\rho_{2,0}$ for various $\Lambda$ values and $\eta/s = 0.2,0.02$.}
    \label{fig:Rho25}
    \end{subfigure}
     \caption{Evolution of the scaled moments and their dependence on \(\Lambda\) and initial specific viscosity $\eta/s$.} 
\end{figure}

In Fig.\eqref{fig:Rho21} we plot the scaled moment $\rho_{2,1}/\rho_{2,0} (=(P_L-P_T)/\epsilon)$ which is related to the pressure anisotropy of the system as a function of scaled time $\Bar{\tau} = T^{-\Lambda}(\tau/\tau_R^0)$. The moments have been plotted for various values of $\Lambda$ and two different sets of initial  $\eta/s$. We see that for larger $\Lambda$, expansion generates larger anisotropies before decaying to zero. The reason is that the momentum-dependent relaxation time taken here scales as the positive exponents of the single-particle energy $E_{p}^{\Lambda}$. So, particles at higher energy (momentum) and with larger $\Lambda$ values are expected to have longer relaxation times and a slower decay rate before restoring the equilibrium. %This allows the expansion to drive the system to larger values of anisotropy and $\rho$ moments with higher magnitude with increasing values $\Lambda$. 
The separation between the peak anisotropy values are also observed to increase with larger momentum dependence.  We further observe that as $\Lambda$  varies, the system with a smaller initial $\eta/s$ shows a larger relative increase in anisotropy and hence a greater sensitivity to $\Lambda$. 

\onecolumngrid

\begin{figure}[!h]
 \begin{subfigure}[!h]{1\textwidth}
    \centering
    \includegraphics[width=0.40\linewidth]{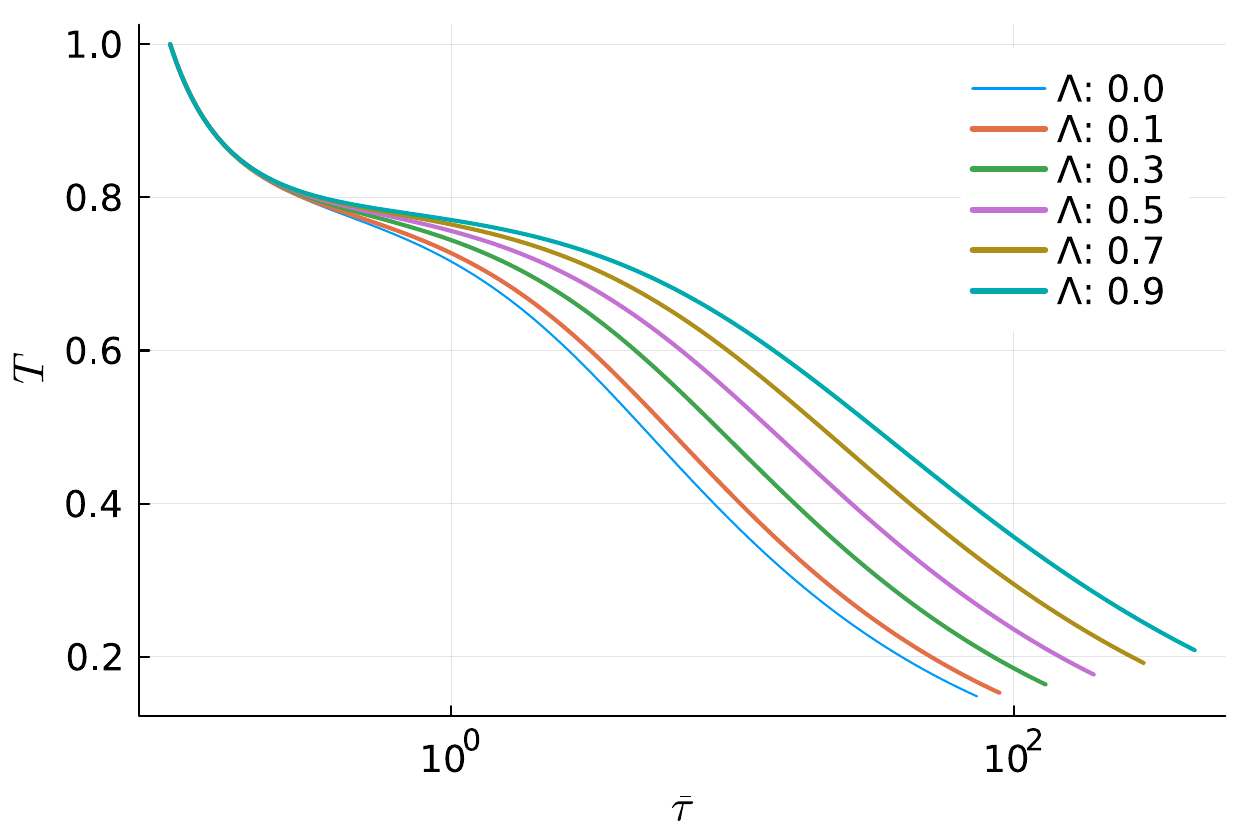} 
    \includegraphics[width=0.40\linewidth]{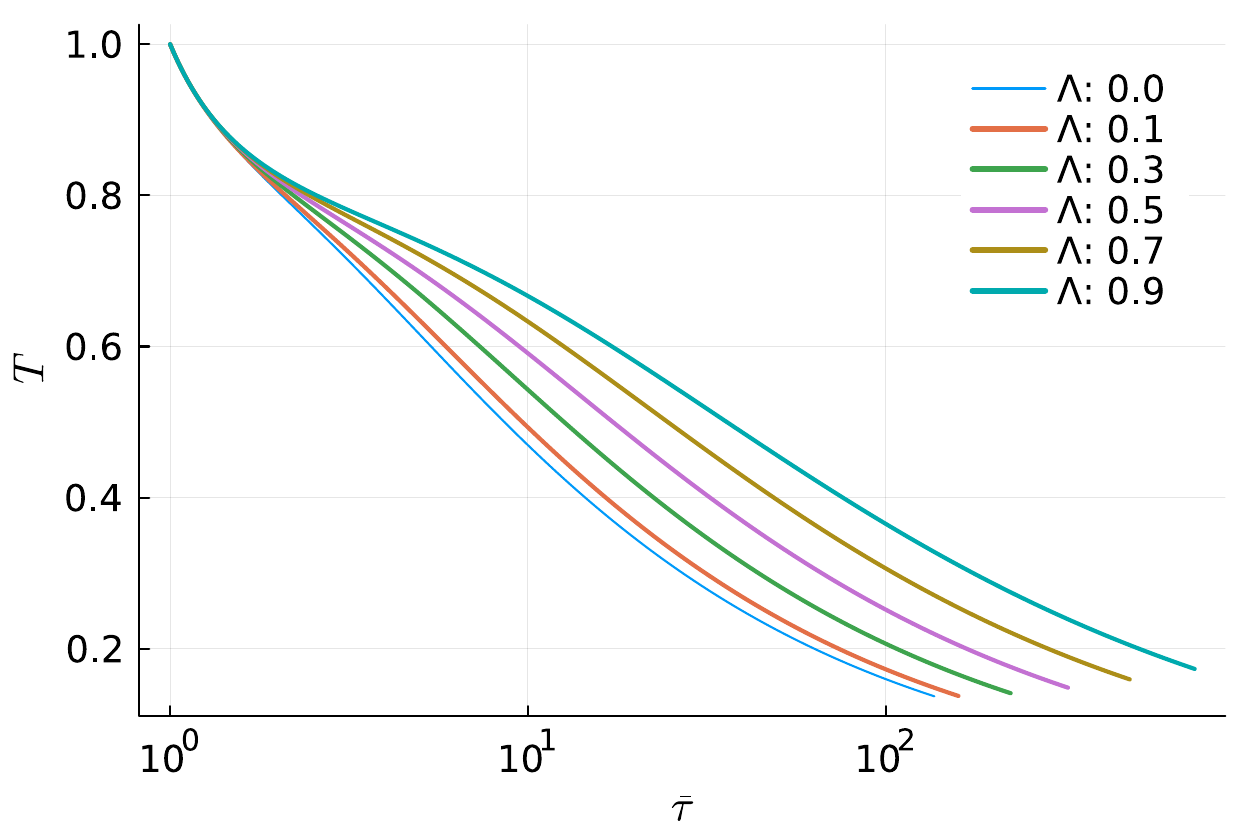} 
    \caption{Evolution of temperature for various $\Lambda$ values and initial $\eta/s = 0.2,0.02$.}
    \label{fig:Temp}
     \end{subfigure}
    \begin{subfigure}[!h]{1\textwidth}
    \centering 
    \includegraphics[width=0.40\linewidth]{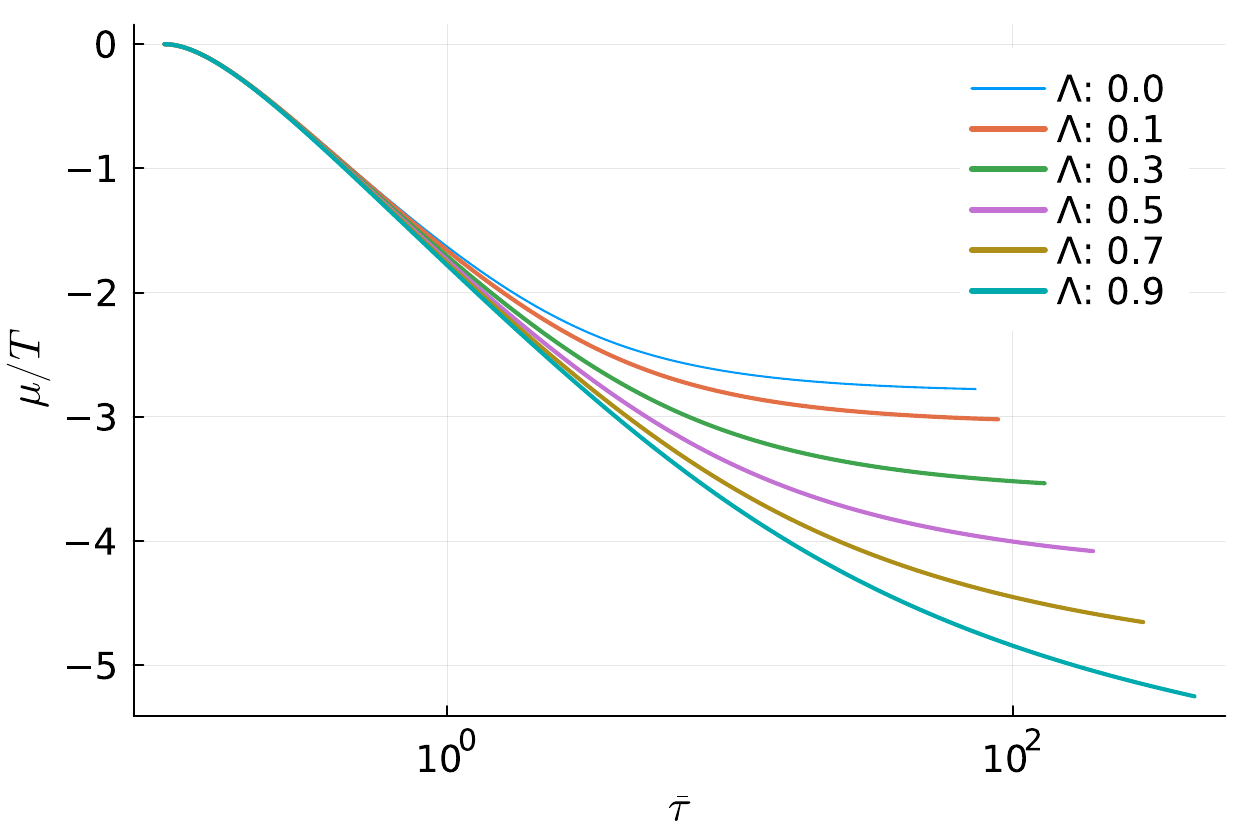} 
    \includegraphics[width=0.40\linewidth]{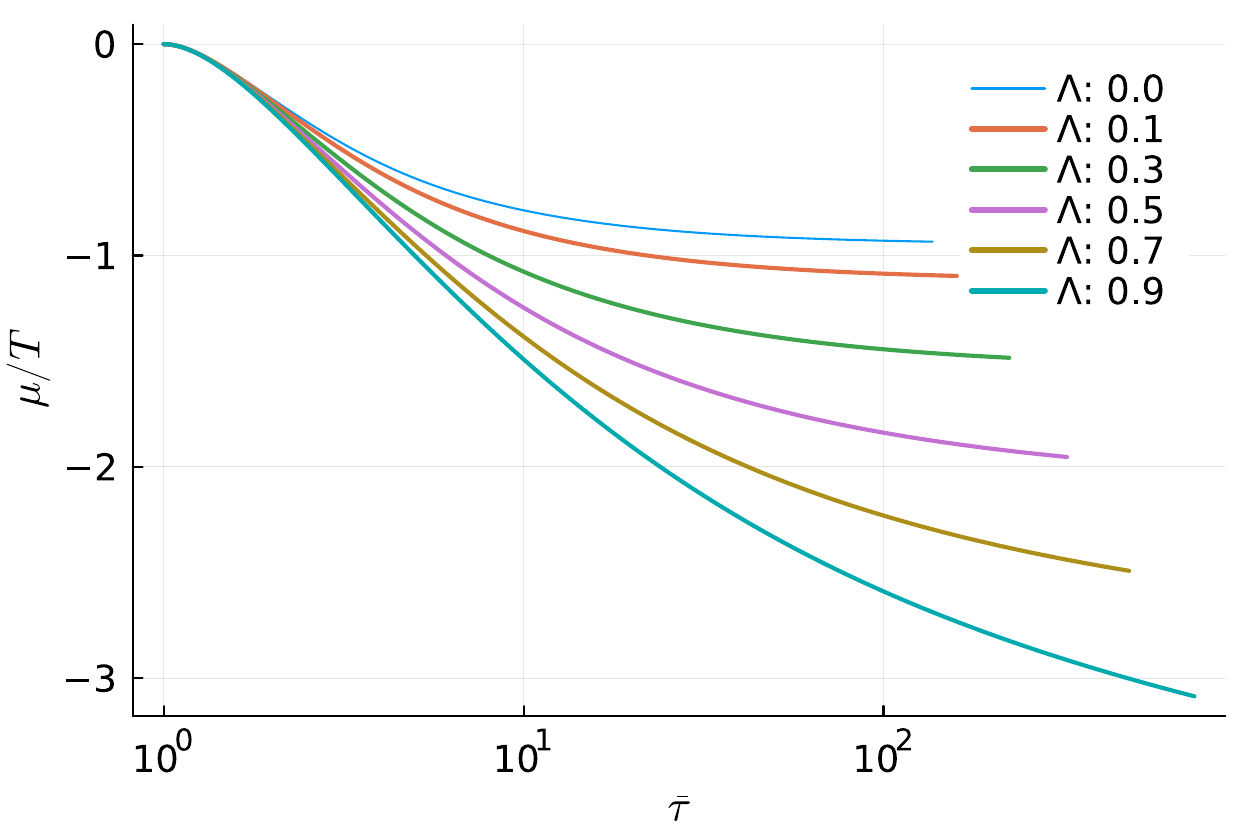}
    \caption{Evolution of chemical potential to temperature ratio for various $\Lambda$ values and initial $\eta/s= 0.2,0.02$.}
    \label{fig:Chem}
    \end{subfigure}
     \caption{Dependence of temperature and \(\mu/T\) evolution on $\Lambda$ and initial specific viscosity $\eta/s$. } 
\end{figure}
\twocolumngrid
\clearpage
 In Fig.\eqref{fig:Rho25} we plot the scaled moment $\rho_{2,5}/\rho_{2,0}$ as a function of $\Bar{\tau}$.  Larger $l$ moments give us the information about the higher angular variation of the out-of-equilibrium distribution function.  Like in the previous $\rho_{2,1}$ case, the peak anisotropy value is higher for increasing $\Lambda$ values. It also shows a larger separation in the peak values when $\Lambda$ is increased. This indicates a sharper sensitivity of large $l$ moments on \(\Lambda\). The sensitivity of $l$ moments on $\Lambda$ is particularly pronounced for lower $\eta/s$.  We see that the decay time of the moment increases for increasing values of $\Lambda$. This implies that the system approaches equilibrium significantly slower when the relaxation time has a larger dependence (larger $\Lambda$) on momentum.

Fig.\eqref{fig:Temp} shows the temperature evolution of the system for various values of $\Lambda$ and initial viscosity. The qualitative features of the temperature evolution remain the same for varying values of $\Lambda$. Initially, there is an expansion-driven cooling. The temperature evolution flattens in the intermediate region, showing a competition between expansion and viscous effects. Finally, the system approaches equilibrium showing a power law decay in the temperature irrespective of $\Lambda$ values, where viscous effects dominate the cooling. For the lower value of initial viscosity, this intermediate region is shorter because of a larger collision rate.  However, we see that for larger values of $\Lambda$, the intermediate phase is longer and therefore a slower approach toward equilibrium is observed, as expected. Fig.\eqref{fig:Chem} shows the evolution of chemical potential to temperature ratio $\mu/T$ for different values of $\Lambda$ and initial $\eta/s$. Like the temperature dependence, here the qualitative features remain the same but it shows a higher negative value with increase in $\Lambda$ indicating larger asymmetry in the system.

\section{CONCLUSIONS AND OUTLOOK}
MDRTA provides a technique to include the microscopic momentum anisotropies of particle interaction in the collision term of the relativistic kinetic equation. Although the momentum dependence of particle interaction is modeled in a simple energy exponent form in the microscopic relaxation time scale, it is certainly an improvement over the constant relaxation time (mostly a thermal average taken) that assumes identical equilibrium restoration times for all particles with different momenta.
In the current study, we employ such a momentum-dependent relaxation time to explore the evolution of the single-particle moments while preserving the Bjorken symmetry. We find that incorporating MDRTA into the method of moments gives rise to considerable analytical complexity but at the same time interesting physical insights describing the momentum anisotropies in a dissipative medium.

Existing studies with boost invariance and momentum-independent RTA show that in the moment hierarchy, the indices corresponding to the ranks of the irreducible momentum tensor (which, under Bjorken symmetry, align with even degrees of the Legendre polynomial), i.e, the \(l\) index in \(\rho_{n,l}\) become coupled.
 However, the indices related to the energy exponents \(n\) do not mix and gives rise to decoupled equations of motion for each \(n\)-th moment. The situation becomes much more involved with MDRTA being introduced in the kinetic equation. The different \(n\) moments now become coupled via the $\Lambda$ parameter (which controls the exponent in the momentum dependence in $\tau_R(p)$) of MDRTA. The coupled differential equations now show atypical behavior of late time growth ($\tau/\tau_R \gg 1$) instead of decaying to zero.  But by setting a tuning parameter, we approximated the behavior for a late time decay.
 
  The following observations are in order: \\
 (i) The temperature evolution as a function of scaled time \(\tau/\tau_{R}\) becomes flatter around \(\tau/\tau_R \sim1\) when collision starts dominating over free streaming with increasing value of $\Lambda$, reflecting a slower cooling when relaxation time depends on particle momenta. \\
 (ii) For fixed orders of \(n\) and \(l\), the magnitude of $\rho$ moments increases with larger $\Lambda$, suggesting enhanced momentum anisotropy in the system.  The system also shows a greater sensitivity on \(\Lambda\) through higher order moments  for smaller \(\eta/s\) and early initial time.\\
  (iii) Increased anisotropy at intermediate times for larger $\Lambda$ implies a later isotropisation and therefore slower approach to local equilibrium.
 
 This behavior is the result of interplay of competing mechanisms. Different angular anisotropies gets coupled through the free streaming dynamics with a \(1/\tau\) dependence. While simultaneously, the collision kernel which depends on temperature and shear viscosity to entropy density ratio (\(\eta/s\)) introduces an admixture of \(n\) moments that act in opposition. The balance between these mechanisms ultimately governs the observed evolution of moments.

To the best of our knowledge, this work represents the first comprehensive investigation utilizing microscopic momentum transfers directly within the kinetic equation governing moment evolution. Our approach provides new understanding of the  dynamics of non-equilibrium systems with momentum dependent relaxation. The analysis opens up the room for some theoretical concepts as well, such as the atypical behavior of late time growth could be traced from the current form of momentum dependence taken in $\tau_R(p)$ which has been adopted from the existing literature \cite{Teaney:2013gca,Rocha:2021zcw}. A more involved momentum dependence in the relaxation rate, especially depending on particle energy scales, could be a line of investigation for future endeavors.

 %We believe that these results hold merit in the study of moment evolution since this is the first attempt to best of our knowledge that utilize the microscopic momentum transfers in the kinetic equation that governs the evolution of the moments.

In future works, we aim to extend the current method to systems with massive particles and fewer symmetry constraints. %Studying the trace anomaly in a massive, non-conformal system using moments could be an interesting and insightful exploration. 
Furthermore, a direct comparison between the moment solutions and the iterative hydrodynamic solutions of macroscopic variables is another direction for future exploration.

\section{ACKNOWLEDGMENTS}
S.M. and V.R. would like to acknowledge the financial support from DAE India through the RIN 4001 project. V.R. is also supported by funding from Anusandhan National Research Foundation (ANRF) CRG/2023/001309, India. We thank the anonymous referee for providing us the proof of positive definiteness of the regularized collision matrix.  
%\clearpage
%%%%---------------------------------------------------------------------------------------------------------------

\appendix
\begin{widetext}
\section{Collision Kernal}\label{Apx:ColKer}

The RTA collision kernel introduced in \cite{Rocha:2021zcw}  is
\begin{equation}\label{Eq:NovCol}
    \hat{L}_{RTA}\phi(p) = -\frac{p}{\tau_R} f^{eq}\lrb{\phi -\frac{\langle p/\tau_R \phi \rangle_{eq} }{\langle p/\tau_R\rangle_{eq} } - P_1 \frac{\langle p/\tau_R\phi P_{1}^{0}\rangle_{eq} }{\langle p/\tau_R P_{1}^{0}P_{1}^{0}\rangle_{eq} }  - 3 p^{\langle \mu\rangle} \frac{\langle p/\tau_R p_{\langle \mu\rangle}\phi \rangle_{eq}}{\langle p/\tau_R p_{\avg{\nu}}p^{\avg{\nu}}\rangle_{eq} } }\,,
\end{equation}
where we have used the massless condition $p^{0} = |\Vec{p}| \equiv p$. Here 
\begin{equation}
    P_{1} = 1 - \frac{\avg{p/\tau_{R}}_{eq}}{\avg{p^2/\tau_{R}}_{eq}}p \,,
\end{equation}
\begin{equation}
    \phi = \frac{f - f^{eq}}{f^{eq}}\,,
\end{equation}
and the angular bracket is defined as
\begin{align}
    \avg{ ... } &= \int dP (...)f \,.\\
    \avg{ ... }_{eq} &= \int dP (...)\feq \,.
\end{align}
For notational convenience, we rewrite the Eq.\eqref{Eq:NovCol} as
\begin{align}
    \hat{L}_{RTA}\phi(k) &= -\frac{p}{\tau_R} f^{eq}\lrb{\phi -A -  B P_1 - C_{\langle \mu\rangle}k^{\langle \mu\rangle}  }~,\\
    & =   -L_1 - L_2 -L_3 -L_4\,,
\end{align}
where $L_{i}$ denote the $4$ terms in the collision kernel. We use a momentum-dependent relaxation time of the form,
\begin{align}
    \frac{1}{\tau_R } = \frac{1}{\tau_R^{0}p^{\Lambda}} ~,
\end{align}
where $\tau_{R}^{0}$ is independent of momentum. The moments of the distribution function is defined as,
\begin{align}
    \rho_{n,l} &= \int dP p^{n}P_{2l}(\cos{\theta}) f (\tau,p_T,p_\eta)~.
\end{align}

We would like to compute here the quantity,
\begin{equation}
  C_{n,l}  = \int dP p^{n}P_{2l}(\cos{\theta}) \hat{L}_{RTA}\phi(k)~.
\end{equation}

%%---------------------------- FIRST TERM -----------------------------------------------

\subsection*{The first term $L_1$}
\begin{align}
     \left< L_1 \right>  &=\int \frac{d^3p}{(2\pi)^{3}p} p^{n}P_{2l}(\cos{\theta}) \lrrb{\frac{p}{\tau_R^{0}} f^{eq} \phi} \,,\nonumber\\
     &= \frac{1}{\tau_R^{0}}\int \frac{d^3p}{(2\pi)^{3}} p^{n-\Lambda}P_{2l}(\cos{\theta})  \lrb{f - f^{eq}}\,, \nonumber\\
     & = \frac{1}{\tau_R^{0}}\lrb{\rho_{n-\Lambda \, l} - \rho_{n-\Lambda\, l}^{eq}}\,.
\end{align}

%%---------------------------- THIRD TERM -----------------------------------------------
\subsection*{The second term $L_2$}

\[
L_2 =  A \frac{p^{1-\Lambda}}{\tau_R} f^{eq} ~,
\]

\begin{align}
     \left< L_2 \right> &= \int \frac{d^3p}{(2\pi)^{3}} p^{n}P_{2l}(\cos{\theta}) \lrrb{\frac{p^{-\Lambda}}{\tau_R} f^{eq}}~,\\
     &=  \frac{1}{{\tau_{R}^{0}}}\rho_{n-\Lambda\, l}^{eq}\delta_{l,0}~.
\end{align}

\textbf{The coefficient A}
\begin{align}
    A &=\frac{\langle p/\tau_R \phi \rangle_{0} }{\langle p/\tau_R \rangle }_{0}~, \\
    &=\frac{\rho_{1-\Lambda,0} - \rho_{1-\Lambda,0}^{eq}}{\rho_{1-\Lambda,0}^{eq}}~.
\end{align}

%%---------------------------- THIRD TERM -----------------------------------------------
\subsection*{The third term $L_3$}

\begin{align}
    \frac{L_3 }{p} &= -B P_{1}^{0} \,,\nonumber\\
    &= -B \lrb{1 - \frac{\avg{p/\tau_{R}}_{eq}}{\avg{p^2/\tau_{R}}_{eq}}p}\,.
\end{align}

Using ,
\begin{align}
    P_{1}^{0} &= 1 - \frac{\avg{p/\tau_{R}}_{0}}{\avg{p^2/\tau_{R}}_{eq}}p ~,\nonumber\\
    &= \lrrb{1 - \frac{\rho_{1-\Lambda,0}^{eq}}{\rho_{2-\Lambda,0}^{eq}} p}\,.
\end{align}

We get,
\begin{align}
    \int dP p^{n}P_{2l}(\cos{\theta}) \lrrb{\frac{p^{-\Lambda}}{\tau_R^0}P_{1}^{0}} &= \frac{1}{\tau_R^{0}}\int dP p^{n-\Lambda}P_{2l}(\cos{\theta})\lrrb{1 - \frac{\rho_{1-\Lambda,0}^{eq}}{\rho_{2-\Lambda,0}^{eq}} p}\,,\nonumber\\
    &= \frac{1}{\tau_R^{0}} \lrb{\rho_{n+1-\Lambda,l}^{eq}- \frac{\rho_{1-\Lambda,0}^{eq}}{\rho_{2-\Lambda,0}^{eq}} \rho_{n+2-\Lambda,l}^{eq} }\delta_{l,0}\,.
\end{align}

The delta function is due to all equilibrium integrals for $l>0$ being zero. The numerator of \textbf{the coefficient $B$} is
\begin{align}
   \text{ Nm}(B) &= \avg {\frac {p}{\tau_R} \phi P_{1}^{0}}_{eq}~,  \nonumber\\
    &= \lrb{\lrb{\rho_{1-\Lambda,0} - \rho_{1-\Lambda,0}^{eq}} - \frac{\rho_{1-\Lambda,0}^{eq}}{\rho_{2-\Lambda,0}^{eq}} \lrb{ \rho_{2-\Lambda,0} - \rho_{2-\Lambda,0}^{eq} }}/\tau_R^{0} ~,\nonumber\\
    &= \lrb{\rho_{1-\Lambda,0} - \frac{\rho_{1-\Lambda,0}^{eq}}{\rho_{2-\Lambda,0}^{eq}} \rho_{2-\Lambda,0} }/{\tau_R^{0}}~.
\end{align}

We have,
\begin{equation}
    \lrb{P_{1}^{0} }^{2}= \lrb{1 - 2\frac{\rho_{1-\Lambda,0}^{eq}}{\rho_{2-\Lambda,0}^{eq}}p + \lrb{\frac{\rho_{1-\Lambda,0}^{eq}}{\rho_{2-\Lambda,0}^{eq}}}^2 p^2}\,.
\end{equation}
The denominator of \textbf{the coefficient $B$} is,
\begin{align}
   \text{Dm}(B) &= \langle p/\tau_R P_{1}^{0}P_{1}^{0}\rangle_{eq}~, \\
    &= \lrb{\rho_{1-\Lambda,0}^{eq} - 2\frac{\rho_{1-\Lambda,0}^{eq}}{\rho_{2-\Lambda,0}^{eq}} \rho_{2-\Lambda,0}^{eq} +    \lrb{\frac{\rho_{1-\Lambda,0}^{eq}}{\rho_{2-\Lambda,0}^{eq}}}^2 \rho_{3-\Lambda,0}^{eq}}/\tau_R^{0}~, \\
    &= \lrb{-\rho_{1-\Lambda,0}^{eq}    + \lrb{\frac{\rho_{1-\Lambda,0}^{eq}}{\rho_{2-\Lambda,0}^{eq}}}^2 \rho_{3-\Lambda,0}^{eq}}/\tau_R^{0} ~.
\end{align}

%%---------------------------- FOURTH TERM -----------------------------------------------
\subsection*{The fourth term }
\begin{align}
    L_4 &=\int dP p^{n+1-\Lambda}P_{2l}(\cos{\theta}) p^{\avg{\mu}} ~,\nonumber\\
        &= \int dP p^{n+1-\Lambda}P_{2l}(\cos{\theta}) p^{\avg{\mu}} ~,\, \nonumber\\
     &= \frac{(2l-1)!!}{l!}z_{\avg{\mu_1\dots\mu_l}}\int dP p^{n+1-\Lambda} p^{\avg{\mu_1\dots\mu_l}}p^{\avg{\mu}}\,.
\end{align}
here we use the identity in \cite{deBrito:2024vhm} (Eq.[33]).
To compute this, we use the identities \cite{deBrito:2024vhm} (Eq.[21b])
\begin{align}
    p^{\avg{\mu_1\dots\mu_l}}p^{\avg{\mu_{l+1}}} = p^{\avg{\mu_1\dots\mu_{l+1}}} + \frac{l}{2l+1}\lrb{\Delta_{\lambda\beta}p^{\lambda}p^{\beta}}\Delta^{\mu_1\dots\mu_{l}}_{\alpha_1\dots\alpha_{l}} \Delta^{\alpha_l\mu_{l+1}} p^{\avg{\alpha_1\dots\alpha_{l-1}}},
\end{align}
\cite{deBrito:2024vhm} (Eq.[17])
\begin{align}\label{Eq:IrrfEq}
   \int dP p^{r}p^{\avg{\mu_1\dots\mu_l}}f^{eq} = 0 && l>0~,
\end{align}
and,
\begin{equation}
    \lrb{\Delta_{\lambda\beta}p^{\lambda}p^{\beta}} = \lrb{(p^{0})^2 - m^2} = -p^2\,.
\end{equation}

Setting $l \to 2l$ for the conformal Bjorken case, we see that the third term contains only odd moments. They are therefore identically zero as $f$ is even in $\theta$.

Finally, putting all the terms together we get the full collision kernel,

\begin{align}
    C_{n,l,\Lambda}=&-\frac{1}{\tau^{0}_{R}} \Bigg[\left\{\rho_{n-\Lambda,l}-\delta_{l0}~\rho^{eq}_{n-\Lambda,l}\right\}
    -A~\delta_{l0}~\rho^{eq}_{n-\Lambda,0}\nonumber\\
    &-B~\delta_{l0}~\left\{\rho^{eq}_{n-\Lambda,0}-\frac{\rho^{eq}_{1-\Lambda,0}}{\rho^{eq}_{2-\Lambda,0}}\rho^{eq}_{n-\Lambda+1,0}\right\}    \Bigg]~,
\end{align}
where
\begin{align}
    A=&\frac{\rho_{1-\Lambda,0}-\rho^{eq}_{1-\Lambda,0}}{\rho^{eq}_{1-\Lambda,0}}~,\\
    B=&\frac{\rho_{1-\Lambda,0}-\frac{\rho^{eq}_{1-\Lambda,0}}{\rho^{eq}_{2-\Lambda,0}}\rho_{2-\Lambda,0}}
    {\rho^{eq}_{3-\Lambda,0}\left(\frac{\rho^{eq}_{1-\Lambda,0}}{\rho^{eq}_{2-\Lambda,0}}\right)^2-\rho^{eq}_{1-\Lambda,0}}~.
\end{align}

Further simplification can be achieved if we use the equilibrium moment expression,
\begin{equation}
     \rho_{n,l}^{eq}(T,\mu)  = e^{\mu/T}\frac{T^{n + 2}}{2\pi^2}\Gamma(n+2)\delta_{l0}\,,
\end{equation}

\begin{align}
    C_{n-1,0} &= - \frac{1}{\tau_R} \rho_{n-\Lambda,0} \lrrb{ 1 - T^{n-1} \frac{\Gamma(n-\Lambda+2)}{\Gamma(1-\Lambda+2)}\frac{\rho_{1-\Lambda,0}}{\rho_{n-\Lambda,0}}}  \nonumber\\
    & + \frac{T^{n-2}}{\tau_R} \rho_{2-\Lambda,0} \lrb{\frac{\Gamma(n-\Lambda+2)}{\Gamma(2-\Lambda+2)}} \lrrb{ 1 - T \frac{\Gamma(2-\Lambda+2)}{\Gamma(1-\Lambda+2)}\frac{\rho_{1-\Lambda,0}}{\rho_{2-\Lambda,0}} } \frac{  \lrrb{ 1 - \frac{\Gamma(1-\Lambda+2)}{\Gamma(2-\Lambda+2)}\frac{\Gamma(n+ 1-\Lambda+2)}{\Gamma(n-\Lambda+2)}    } }{ \lrrb{1 -  \lrrb{\frac{\Gamma(1-\Lambda+2)}{\Gamma(2-\Lambda+2)}}^2\frac{\Gamma(3-\Lambda+2)}{\Gamma(1-\Lambda+2)}  }}\,.
\end{align}
By defining the two functions, 
\begin{align}
    & K(n,m,\Lambda)= \frac{\Gamma(n-\Lambda +2)}{\Gamma(m-\Lambda +2)} ~,\\
    & C(n,\Lambda)= \frac{1 - K(1,n,\Lambda)K(n+1,2,\Lambda)}{1 - K(1,2,\Lambda)K(3,2,\Lambda) }~,
\end{align}
we get the simplified expression,
\begin{align}
    C_{n,l,\Lambda}= &-\frac{1}{\tau_R} \Big[\rho_{n-\Lambda,l}\nonumber\\
    &- \delta_{l0}\Big\{ T^{n-1}K(n,1,\Lambda)[1 - C(n,\Lambda)]\rho_{1-\Lambda,l} \nonumber\\
    &~~~~~~+ T^{n-2}K(n,2,\Lambda)C(n,\Lambda)\rho_{2-\Lambda,l}\Big\}\Big]~.
\end{align}

Note that
\begin{align}
    K(n,n,\Lambda) &= \frac{\Gamma(n-\Lambda +2)}{\Gamma(n-\Lambda +2)} = 1\,,\\
    C(1,\Lambda) &=\frac{1 - K(1,1,\Lambda)K(2,2,\Lambda)}{1 - K(1,2,\Lambda)K(3,2,\Lambda) } = 0\,\\
    C(2,\Lambda) &=\frac{1 - K(1,2,\Lambda)K(3,2,\Lambda)}{1 - K(1,2,\Lambda)K(3,2,\Lambda) }= 1\,,
\end{align}

and enforce the conservation laws for energy density $\epsilon = \rho_{2,0}$ and number density $n = \rho_{1,0}$,
\begin{align}
    C_{1,0,\Lambda} &= 0\,,\nonumber\\
    C_{2,0,\Lambda} &= 0\,,\nonumber
\end{align}
independent of the value of $\Lambda$.

\newpage
\section{Divergence and Closure}

\subsection{Vector representation of coupled equations}\label{Ap:Vec}

We used the expression \eqref{Eq:MomMat},
 \begin{equation}\label{Eq:MomMat2}
     \dv{\vec{\rho}}{\tau}= -\frac{1}{\tau} \mathbf{F} \vec{\rho} -\frac{1}{\tau^{0}_R}{\mathbf{C}}(\Lambda)\vec{\rho}\,,
 \end{equation}
to represent the infinite hierarchy of moments and its finite truncations. 
 
We show the vector representation for any finite truncation in $n$ and $l$ for the infinitely coupled differential equations in \eqref{Eq:MomMat}.  Any finite set of linear (semi-linear) coupled differential equations can be written in a matrix form. Considering both the $n$ and $l$ coupling makes the matrices far more complicated as we require an ordering of the $(n,l)$ indices. A simple ordering for $\Lambda = 1$  and for a truncation $n_{max}=3 $, $l_{max} = 3$ is,
 \begin{equation}
     \vec{\rho} = \begin{bmatrix}
         \rho_{0,0} \sim \rho_{0}\\
         \rho_{1,0}\sim \rho_{1}\\
         \rho_{2,0}\sim \rho_{2}\\
         \rho_{0,1}\sim \rho_{3}\\
         \rho_{1,1}\sim \rho_{4}\\
         \rho_{2,1}\sim \rho_{5}\\
         \rho_{0,2}\sim \rho_{6}\\         
         \rho_{1,2}\sim \rho_{7}\\
         \rho_{2,2}\sim \rho_{8}\\
     \end{bmatrix} \label{Eq:Vector}\,.
 \end{equation}
 In this form we can forgo the $(n,l)$ indexing and choose a single index, say, $m$. We note that  $\mathbf{C}(\Lambda)$  only couples the $n$ and $\mathbf{F}$ only couples the $l$. This implies we can always rearrange the matrices in a form where either $\mathbf{C}(\Lambda)$ or $\mathbf{F}$ is box diagonal. The ordering we used in \eqref{Eq:Vector} is such that it makes the matrix $\mathbf{C}(\Lambda)$ box diagonal. We explicitly show these two matrices below for the vector \eqref{Eq:Vector},
 \begin{equation}
   \mathbf{F} = \begin{bmatrix}
       Q_{0,0}  & 0         &0          & R_{0,0}   & 0         & 0         & 0         & 0         &0      \\
       0        & Q_{1,0}   &0          & 0         & R_{1,0}   & 0         & 0         & 0         &0      \\
       0        & 0         &Q_{2,0}    & 0         & 0         & R_{2,0}   & 0         & 0         &0      \\
       P_{0,1}  & 0         &0          & Q_{0,1}   & 0         & 0         & R_{0,1}   & 0         &0      \\
       0        & P_{1,1}   &0          & 0         & Q_{1,1}   & 0         & 0         & R_{1,1}   &0      \\
       0        & 0         &P_{2,1}    & 0         & 0         & Q_{2,1}   & 0         & 0         &R_{2,1}\\
       0        & 0         &0          & P_{0,2}   & 0         & 0         & Q_{0,2}   & 0         &0      \\
       0        & 0         &0          & 0         & P_{1,2}   & 0         & 0         & Q_{1,2}   &0      \\
       0        & 0         &0          & 0         & 0         & P_{2,2}    & 0         & 0         &Q_{2,2}      
   \end{bmatrix}  \,,
 \end{equation}
 and
 \begin{equation}
   \mathbf{C(\Lambda = 1)} = \begin{bmatrix}
       0                & 0                 &0                  & 0                 & 0                 & 0                 & 0         & 0         &0      \\
       c_{1-\Lambda,0}  & 0                 &0                  & 0                 & 0                 & 0                 & 0         & 0         &0      \\
       0                & c_{2-\Lambda,0}   &0                  & 0                 & 0                 & 0                 & 0         & 0         &0      \\
       0                & 0                 &0                  & 0                 & 0                 & 0                 & 0         & 0         &0      \\
       0                & 0                 &c_{1-\Lambda,1}    & 0                 & 0                 & 0                 & 0         & 0         &0      \\
       0                & 0                 &0                  & c_{2-\Lambda,1}   & 0                 & 0                 & 0         & 0         &0      \\
       0                & 0                 &0                  & 0                 & 0                 & 0                 & 0         & 0         &0      \\
       0                & 0                 &0                  & 0                 & c_{1-\Lambda,2}   & 0                 & 0         & 0         &0      \\
       0                & 0                 &0                  & 0                 & 0                 & c_{2-\Lambda,2}   & 0         & 0         &0      
   \end{bmatrix}  \,.
 \end{equation}

 In $\mathbf{C(\Lambda)} $ we gave above, the coupling of $l=0$ moments to energy and number density moments have not been shown. But it should be clear that the coupling between $n$ and $l$ moments is taken care of by equation \eqref{Eq:MomMat2} and can always be written in the single index form in \eqref{Eq:Vector}.

\subsection{Moment Closure}\label{Apx:MomCls}

We observed that the numerical solution to a naive truncation of the moment equations does give a positive definite collision kernel. We remedy this by using the Grad closure procedure, making use of the momentum expansion of the distribution function. We assume that the distribution function can be expanded in momentum as
\begin{equation}\label{Eq:GradMomExp}
    f(\tau,p,\cos{\lrb{\theta}}) = f^{\text{eq}}\sum_{m=0}^{N}\sum_{k=0}^{L}A_{m,k}(\tau)p^{m}P_{2k}(\cos{\theta}).
\end{equation}

The moments $\rho_{n,l}$, then can be expressed as
\begin{align}
    \rho_{n+\Lambda,l} &= \sum_{m=0}^{N}\sum_{k=0}^{L}A_{m,k}\frac{1}{(2\pi)^3}\int d\theta\sin{\theta} d\phi dp~p^{n+m+1 +\Lambda}P_{2l}(\cos{\theta})P_{2k}(\cos{\theta})f^{\text{eq}} \,,\nonumber\\
    &= \sum_{m=0}^{N}\sum_{k=0}^{L}A_{m,k} \delta_{kl}\frac{e^{\alpha}}{2\pi^2}\frac{T^{n+m+2+\Lambda}}{4k+1}\Gamma(m+n+2+\Lambda) \,,
\end{align}

To simplify the expressions, we scale the coefficients by,
\begin{equation}
    A_{m,l} \frac{e^{\alpha}}{2\pi^2}\frac{1}{4l+1} \to A_{m,l} \,.
\end{equation}

This reduces the above equation to the form,
\begin{align}\label{Eq:MomCofSimp}
    \rho_{n+\Lambda,l} =\sum_{m=0}^{N}A_{m,l} T^{n+m+2+\Lambda}\Gamma(m+n+2+\Lambda)\,.
\end{align}
Note that $\rho_{n,l}$ is related to $A_{n,l}$ of the same $l$. So at each $l$ we get a set of $N$ linear equations relating the moments to the coefficients $A_{m,k}$. To make this explicit, we write the above equation in a matrix form, using the following definitions,

We define a Hankel matrix $\textbf{H}$,  vectors $\Vec{A}_l$ and $\Vec{\rho}_l$ as
\begin{align}
    [\textbf{H}(T,{\Lambda})]_{n,m} &= T^{n+m+2+\Lambda}\Gamma(n + m + 2+\Lambda), \\
    [\Vec{A}_l]_{n} &= A_{n,l},\\
    [\Vec{\rho}_l ]_{n} & = \rho_{n,l}.
\end{align}

Then,
\begin{equation}
   [\Vec{ \rho_{l}}]_{n+ \Lambda} = \sum_{m}[\textbf{H}(T,{\Lambda})]_{n,m} [\Vec{A}_l]_{m},
\end{equation}

or 
\begin{equation} \label{Eq:MomCof}
   \Vec{\rho}_l(\Lambda) =\textbf{H}(T,{\Lambda}) \Vec{A}_l.
\end{equation}

We can first compute the coefficients from the integer moments  $\rho_{n,l}(\Lambda = 0)$ by inverting Eq.\eqref{Eq:MomCof} to get
\begin{equation}
   \Vec{A}_l= \textbf{H}^{-1}(T,0) \Vec{\rho}_l \,.
\end{equation}

Then from Eq.\eqref{Eq:MomCofSimp} we get,
\begin{align}
    \Vec{\rho}_l(-\Lambda) = \, \textbf{H}(T,-\Lambda) \textbf{H}^{-1}(T,{0}) \Vec{\rho}_l(0).
\end{align}

We note that the matrix $\textbf{H}(T,\Lambda)$ has the decomposition
\begin{equation}
    \text{H}(T,{\Lambda}) = T^{2+\Lambda} ~ \text{D}(T)\text{H}(1,{\Lambda})\text{D}(T),
\end{equation}
where $\text{D}(T) = diag(1,T,T^2,\dots,T^{N})$. So to analyze the features of the matrix $\text{M}$,
\begin{align}
    \textbf{M}(T,\Lambda)  &= \textbf{H}(T,-\Lambda) \textbf{H}^{-1}(T,0),\\
                        &= T^{\Lambda}D(T)\textbf{H}(1,-\Lambda) \textbf{H}^{-1}(1,0)D^{-1}(T).
\end{align}
We only need to be concerned with the matrix $\textbf{M}(1,\Lambda) = \textbf{H}(1,-\Lambda) \textbf{H}^{-1}(1,0)$ which is independent of T. Note that the diagonal matrices $D(T)$ simply effects a similarity transformation on $\textbf{M}(1,\Lambda)$ and therefore do not affect the eigenvalues. As this is a fixed matrix independent of the dynamics, one can verify the positive definiteness by numerically computing its eigenvalues. 

For $N = 5$, $\text{eigenvalues}(\textbf{M}(1,0.5)) \in \left\{ 0.24619,~
 0.3122,~
 0.4046,~
 0.5556,~
 0.8614,~
 1.8523\right\}$. In Fig. \eqref{fig:eigminLam} we plot the minimum eigenvalue of $M(1,\Lambda)$ for $N=5$ for varying values of $\Lambda$. We see that the eigenvalue is positive but exponentially decays to zero as $\Lambda$ increases.
\begin{figure}[!h]
    \centering
    \includegraphics[width=0.6\linewidth]{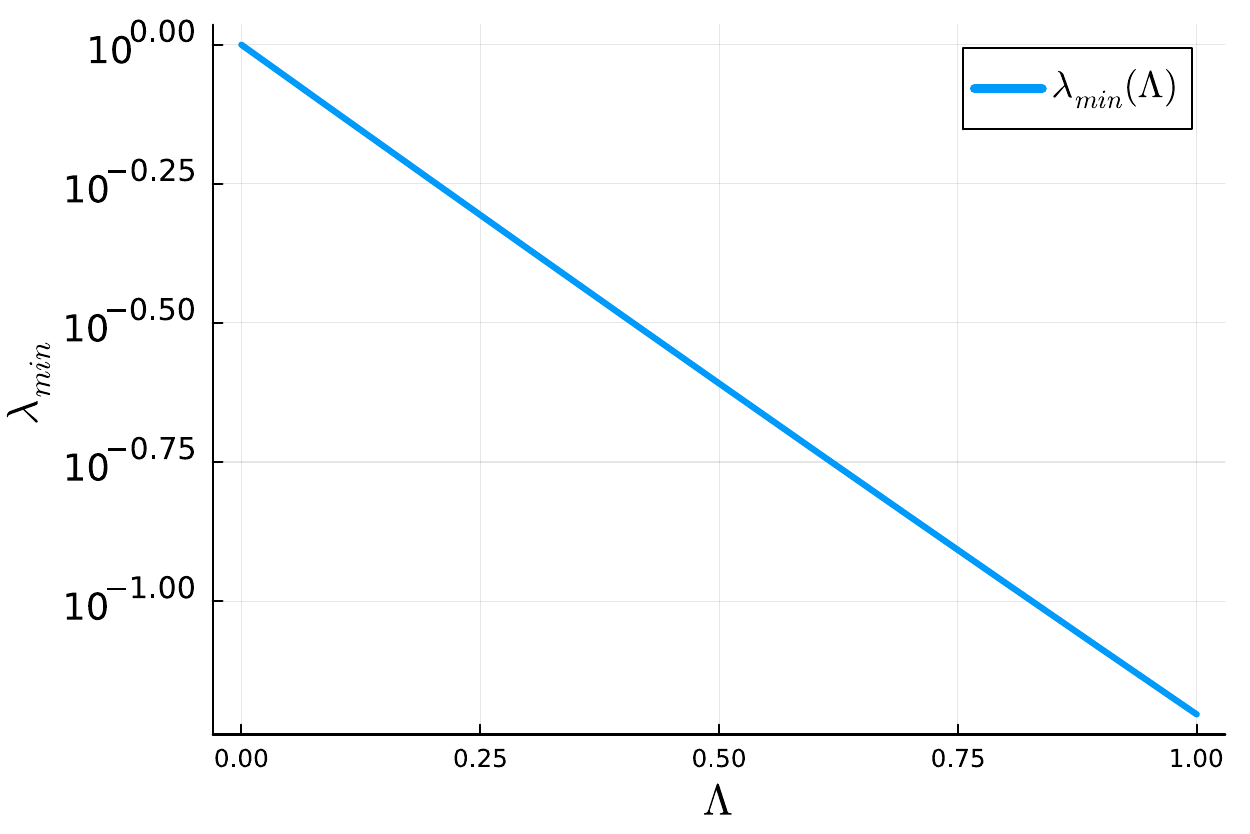}
    \caption{log plot of the minimum eigenvalue of $M(1,\Lambda)$ for $N=5$ as a function of $\Lambda$}
    \label{fig:eigminLam}
\end{figure}

To show this analytically, first we prove that $\textbf{H}(1,\Lambda)$ is positive definite. Consider the quadratic form $x^{T}\textbf{H}(1,\Lambda)x$,
\begin{align}
    x^{T}\textbf{H}(1,\Lambda)x  & = \sum_{n=0}^{N}\sum_{m=0}^{N}~x_{m}x_{n}\Gamma(n+m +2 + \Lambda)\nonumber\\
                                 &=  \sum_{n=0}^{N}\sum_{m=0}^{N}~x_{m}x_{n} \int_{0}^{\infty}dt~ t^{n+m+1 + \Lambda}e^{-t}\nonumber\\
                                 &=  \int_{0}^{\infty}dt~ t^{\Lambda}e^{-t}  \sum_{n=0}^{N}\sum_{m=0}^{N}~x_{m}x_{n}t^{n+m+2} \nonumber\\
                                 &= \int_{0}^{\infty}dt~ t^{\Lambda}e^{-t} \lrb{\sum_{n=0}^{N} x_{n}t^{n+1}}^2 > 0 \,,
\end{align}
where we have used the definition of the gamma function. $\textbf{H}(1,\Lambda)$ is then symmetric and positive definite, which implies that its eigenvalues are positive for all values of $\Lambda$. 
Now the product $\textbf{M}(1,\Lambda) = \textbf{H}(1,\Lambda)\textbf{H}(1,0)$ can be shown to be positive definite by showing that it is similar to a positive definite matrix. For this, multiply both sides of $\textbf{M}(1,\Lambda)$ by $\sqrt{\textbf{H}^{-1}(1,0)}$, the principal square root of $\textbf{H}^{-1}(1,0)$. This gives 
\begin{equation}
    \sqrt{\textbf{H}^{-1}(1,0)}\textbf{M}(1,\Lambda)\sqrt{\textbf{H}^{-1}(1,0)} = \sqrt{\textbf{H}^{-1}(1,0)} \textbf{H}(1,\Lambda)\sqrt{\textbf{H}^{-1}(1,0)} \,.
\end{equation}
Clearly, the new matrix is symmetric. 

Now, for two symmetric positive definite matrices $A$ and $B$, we can show that $A BA$ is also positive definite. Since positive definite matrices are symmetric and invertible, both $A$ and $B$ satisfy $A^T = A$, $B^T = B$, and $x^T A x > 0$, $x^T B x > 0$ for all nonzero vectors $x$. Consider the matrix $C = A B A$. Then
\[
C^T = (A B A)^T = A^T B^T A^T = A B A = C,
\]
so $C$ is symmetric. For any nonzero vector $x$, define $y = A x$. Because $A$ is invertible, $y \neq 0$. We then have
\[
x^T C x = x^T A B A x = (A x)^T B (A x) = y^T B y.
\]
Since $B$ is positive definite, $y^T B y > 0$ for all $y \neq 0$. Therefore $x^T C x > 0$ for all $x \neq 0$, which shows that $C = A B A$ is positive definite. This completes out proof.

As a final note, as $H(1,\Lambda)$ contains factorials as its components, it is an ill-conditioned matrix and inversion becomes numerically inaccurate as $N$ is increased. Therefore, finding accurate $M(1,\Lambda)$ becomes computationally non-trivial for large values of $N$.

\end{widetext}

%\cite{Denicol:2016bjh}
%\cite{Gangadharan:2023yzb}
%\cite{Gangadharan:2024ovs}Romatschke:2009im
\bibliography{main}

\end{document}